\documentclass[12pt]{iopart}

\pdfoutput=1

\usepackage{iopams}  

\usepackage{slashbox, multirow} 
\usepackage{graphicx}
\begin{document}

\title{
Ising exchange interaction in lanthanides and actinides
}

%\author{Content \& Services Team}
%\address{IOP Publishing, Temple Circus, Temple Way, Bristol BS1 6HG, UK}
%\ead{submissions@iop.org}
%\vspace{10pt}
%\begin{indented}
%\item[]February 2014
%\end{indented}

\author{Liviu F. Chibotaru and Naoya Iwahara}
\address{Theory of Nanomaterials Group, 
Katholieke Universiteit Leuven, 
Celestijnenlaan 200F, B-3001 Leuven, Belgium}
\vspace{10pt}
\begin{indented}
\item[]July 29 2015
\end{indented}

\begin{abstract}
The Ising exchange interaction is a limiting case of strong exchange anisotropy and represents a key property of many magnetic materials. 
Here we find necessary and sufficient conditions to achieve Ising exchange interaction for metal sites with unquenched orbital moments. 
Contrary to current views, the rules established here narrow much the range of lanthanide and actinide ions which can exhibit Ising exchange interaction.
It is shown that the Ising interaction can be of two types: (i) coaxial, with magnetic moments directed along the anisotropy axes on the metal sites and (ii) non-coaxial, with arbitrary orientation of one of magnetic moments. 
These findings will contribute to purposeful design of lanthanide and actinide based materials.
\end{abstract}

\vspace{2pc}
\noindent{\it Keywords}: Ising exchange, lanthanides, actinides
%\submitto{\NJP}
%\maketitle

\section{Introduction}
\label{Sec:Introduction}
Strong magnetic anisotropy on the metal sites gives rise to novel magnetic properties, 
such as single-molecule magnet behavior \cite{Gatteschi2006, LayfieldMurugesu2015}, 
magnetic multipole ordering \cite{Zvezdin1985, Santini2009},
and various exotic electronic phases \cite{Witczak-Krempa2014, Gingras2014}. 
In lanthanides and actinides the spin-orbit coupling exceeds the crystal-field splitting of the ground ionic $LS$ term leading to unquenched orbital momentum ${\mathbf{L}}$ \cite{Wybourne1965}. As a result the low-lying spectrum on these metal ions  is well described as crystal-field split eigenstates of the total angular momentum ${\mathbf{J}}={\mathbf{L}}+{\mathbf{S}}$, where ${\mathbf{S}}$ is the spin of the corresponding term. If the symmetry of metallic sites is lower than cubic, the ground crystal-field multiplet is either a Kramers doublet or a quasi doublet, for odd and even number of electrons on the metal ion, respectively. The presence of unquenched orbital momentum in these doublets makes them strongly anisotropic, which is reflected also in the strong anisotropy of their exchange interaction. 
The limit for this anisotropy is the exchange interaction of Ising type, which in the case of two interacting doublets or interacting doublet and isotropic spin has the form:
\begin{eqnarray}
 \hat{\rm H}_{\rm I} &=&  \mathcal{J} \tilde{s}_{1z_1} \tilde{s}_{2z_2}, \nonumber\\
 \hat{\rm H}_{\rm I} &=&  \mathcal{J} \tilde{s}_{1z_1} {\rm S}_{2z_1}, 
\label{Eq:H_coax}
\end{eqnarray}
where $\tilde{s}_{i}$ is the pseudospin 1/2 describing the doublet state of site $i$, ${\rm S}_{2}$ is the isotropic spin of site 2 and $z_i$ denotes the projection of pseudospin or isotropic spin along the main magnetic axis of doublet on site $i$.  

The common beliefs concerning the Ising exchange coupling in lanthanides and actinides can be summarized in the following rules:

1). The interaction involving quasi doublets of metal ions with even number of electrons (the so-called Ising ions \cite{Zvezdin1985}) will always be of Ising type. However, in the case of metal ions with odd number of electrons the Ising interaction is only achieved when the corresponding Kramers doublets have zero magnetic moment in the transversal directions with respect to the main magnetic axis (i.e., are perfectly axial \cite{Ungur2011}) \cite{Zvezdin1985}.

\begin{figure}[bt]
\begin{center}
 \includegraphics[width=7cm, bb=0 0 831 675]{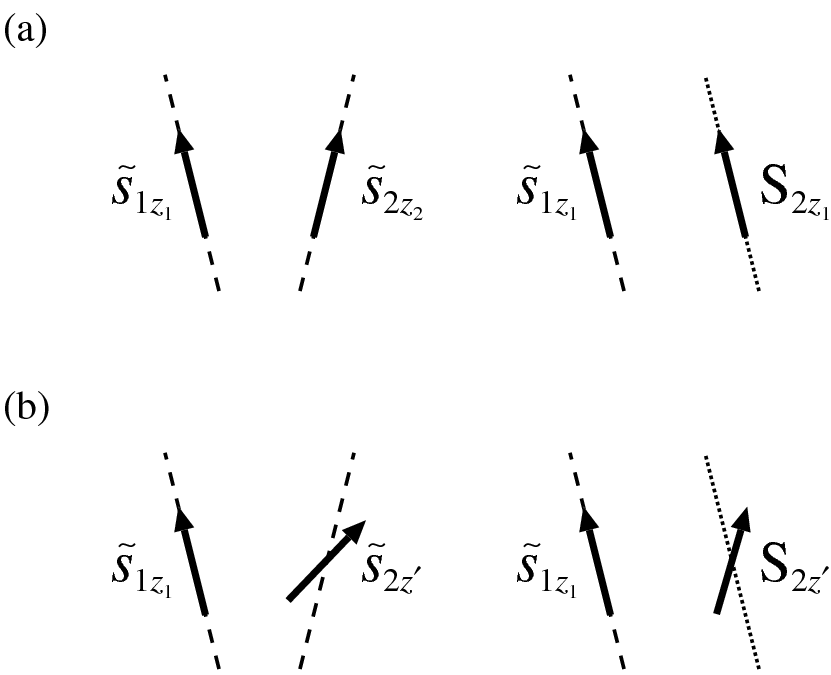}
\end{center}
\caption{ 
Two types of arrangement of local magnetic moments by the Ising exchange interaction: type I - coaxial (a) and type II - non-coaxial (b) . The main magnetic axes on metal sites (dashed lines) are generally non-coplanar. Dotted lines indicate the main magnetic axis of the neighbor site. 
}
\label{Fig1}
\end{figure}

2). In both above cases the Ising interaction is of {\em coaxial} type (\ref{Eq:H_coax}), when the local magnetic moments in the exchange states are directed either along the main magnetic axes $z_i$ of the corresponding metal sites or, in the case of isotropic spin, along the main magnetic axis of the neighbor site (Fig. \ref{Fig1}(a)) \cite{Chibotaru2015}.

Rule 1) appears to be weakly restrictive predicting, in particular, Ising exchange interaction to all metal ions with even number of electrons. As discussed below, both these rules are based on an oversimplified treatment of exchange interaction for metal ions with unquenched orbital momentum. Given the increasing interest for strongly anisotropic magnetic materials, the knowledge of precise conditions for the realization of Ising exchange interaction between strongly anisotropic metal ions would be highly desirable.  

In this work, the general conditions for achieving Ising exchange interaction for metal sites with unquenched orbital momentum are found via analysis of microscopic exchange Hamiltonian. These conditions are specified for interacting anisotropic metal ions in their doublet states and for such ions interacting with isotropic spins. Despite the overwhelming complexity of exchange interaction between centers with unquenched orbital moments, the realization of Ising exchange interaction is shown to depend in most cases solely on the structure of doublet wave functions on the metal sites and not on the details of intersite interaction. This reduces the task of designing magnetic materials with Ising exchange interaction to creation of appropriate crystal field on the metal sites.

\section{Exchange interaction for $J$-multiplets}
\label{Sec:Exchange_J}
The ground-state (quasi) doublet wave functions in lanthanides and actinides can be written as linear combination of eigenstates of the corresponding atomic $J$-multiplet:
\begin{eqnarray}
&&|M\rangle = \sum_{m =-J}^{J}C_{m}|Jm \rangle , \nonumber\\
&&|-M\rangle = \theta |M\rangle,
\label{Eq:doublet}
\end{eqnarray}
where $\theta$ is the operator of time inversion \cite{Abragam1970}. The composition of these wave functions is sensitive to the details of the local crystal field and can involve, in particular, all eigenfunctions $|Jm \rangle$. This means that an adequate description of exchange interaction for these doublet states should involve an interacting Hamiltonian acting on the entire 
%\sout{
%$J$-manifold,
%}
ground $J$-multiplet,
i.e., expressed in terms of total angular momentum operator, ${\rm J}_{\alpha}$, 
$\alpha =x,y,z$. The latter was widely supposed to be of the Heisenberg-like form \cite{Zvezdin1985}:   
\begin{eqnarray}
 \hat{\rm H}_{\rm Heis} = \mathcal{J} {\mathbf{J}}_1 \cdot {\mathbf{J}}_2 . 
\label{Eq:JJ}
\end{eqnarray}
Despite the lack of justification, this form is often used for the description of interaction between lanthanides or actinides (or a similar form, $\propto {\mathbf{J}}_1\cdot {\mathbf{S}}_2$, in the case of their interaction with an isotropic spin) \cite{Molavian2007, Talbayev2008, Curnoe2008, 
Magnani2010, Carretta2013, 
Klokishner2009,  
Dreiser2012, Kofu2013}.
Direct calculations show that with this form of exchange interaction the rules 1-2) hold true 
%\cite{comment}. 
\footnote{The same is true for the phenomenological Lines model \cite{Lines1971}, 
$\hat{\rm H}_{\rm Lines} = \mathcal{J} {\mathbf{S}}_1 \cdot {\mathbf{S}}_2$,  
where ${\mathbf{S}}_1$ and ${\mathbf{S}}_2$ are the spins of the ground state terms on the 
corresponding sites.}.
This is easily seen when we choose a particular form of wave functions (\ref{Eq:doublet}) for strongly axial doublets:
\begin{eqnarray}
&&|\pm M_i \rangle =|J_i \pm m_i \rangle ,\;\;\; i=1,2,
\label{Eq:axial}
\end{eqnarray}
achievable either at high axial symmetry \cite{Ungur2011} or at strong axial component of the crystal field on sites \cite{Blagg2013}. Indeed, it can be checked directly that for odd numbers of electrons the transverse components of any angular momentum is zero for $m_i >1/2$, while in the case of even number of electrons the smallest $m_i$ for a quasi doublet (\ref{Eq:axial}) is $m_i =1$ (the same for Eq. (\ref{Eq:doublet})) . Under such circumstances the rules 1-2) and the conditions for zero transverse magnetization on metal ions (perfect axiality) hold true simultaneously %\cite{comment_1}.
\footnote{A simplified treatment of kinetic exchange interaction 
within so-called $1/U$ approximation \cite{Elliott1968} 
gives for two interacting doublets (\ref{Eq:axial}) on Dy$^{3+}$ ions an 
Ising exchange interaction similar to the first Eq. (\ref{Eq:H_coax}). 
At the same time the interaction of one such Dy$^{3+}$ doublet with an 
isotropic spin results in an Ising exchange, as in the second equation of 
(\ref{Eq:H_coax}), only for $m_1 =J_1 =15/2$ (unpublished results).
}.

One should note that magnetic dipolar interaction between magnetic sites is also bilinear in total angular momentum operators albeit has a form different from Eq. (\ref{Eq:JJ}):
\begin{eqnarray}
 \hat{\rm H}_{\rm dip} &=& -\frac{g_{\rm L}^2 \mu_{\rm B}^2}{R_{12}^3}
\left[{\mathbf{J}}_1 \cdot {\mathbf{J}}_2 
-3 ({\mathbf{J}}_1 \cdot \mathbf{n}_{12}) ({\mathbf{J}}_2 \cdot \mathbf{n}_{12})\right],
\end{eqnarray}
where $g_{\rm L}$ is the Land\'{e} g factor, $\mu_{\rm B}$ is Bohr magneton,
$R_{12}$ is the distance between magnetic sites, and $\mathbf{n}_{12}$ is the normalized direction vector from site 1 to site 2.
Therefore, it will reduce to Eq. (\ref{Eq:H_coax}) too when downfolded on the axial doublets (\ref{Eq:axial}) ($m_i >1/2$). The corresponding constant $\mathcal{J}_{\rm dip}$ in (\ref{Eq:H_coax}) is a simple expression of relative orientations of local main magnetic axes and the vector connecting magnetic sites \cite{Chibotaru2015}. This interaction is considered dominant for pairs of lanthanides ions, which justifies the description of their interactions via bilinear operators (\ref{Eq:JJ}) \cite{Molavian2007, Talbayev2008, Curnoe2008, 
Magnani2010, Carretta2013, Klokishner2009, Dreiser2012, Kofu2013}, as well as the validity of rules 1-2). Such a view is, however, ungrounded for simple reasons that (i) dipolar magnetic coupling can be diminished arbitrarily on geometric grounds and (ii) the exchange interaction can significantly exceed the dipolar coupling, especially, when it is mediated by strongly covalent bridging groups \cite{Tuna2012}. In the cases of lanthanides interacting with isotropic spins (transition metals or radicals) and of interacting actinides (between themselves or with isotropic spins), the magnetic dipolar interaction is significantly smaller than the exchange interaction. The latter, however, is not described by the simple form (\ref{Eq:JJ}) but represents a much more involved expression. In the case of two interacting ions with unquenched orbital momenta it has the form \cite{Santini2009, Iwahara2015}:
\begin{equation}
 \hat{\rm H} = 
 \sum_{kqk'q'} 
  \mathcal{J}_{kqk'q'} 
 \frac{O_k^{q}({\mathbf{J}}_1) O_{k'}^{q'}({\mathbf{J}}_2)}{O_k^0(J_1)O_{k'}^0(J_2)},
\label{Eq:HJJ}
\end{equation}
where 
$\mathcal{J}_{kqk'q'}$ is the exchange parameter, 
$O_k^q({\mathbf{J}}_1)$ and $O_{k'}^{q'}({\mathbf{J}}_2)$ are Stevens operators
\footnote{
The Stevens operator is defined so that it transforms as spherical harmonics under rotation,
which enables us to apply Wigner-Eckart theorem (see e.g., \cite{Chibotaru2012}).
}
whose ranks $k$ and $k'$ have to obey the relation $k+k'=$ even due to the invariance of 
the Hamiltonian with respect to time inversion 
\cite{Abragam1970}. 
In Eq. (\ref{Eq:HJJ}), the terms such that one of $k$ and $k'$ is zero are not included 
because they are not exchange but crystal field.
A metal ion with unquenched orbital momentum interacts with an isotropic spin as follows \cite{Iwahara2015}:
\begin{eqnarray}
 \hat{\rm H} &=& 
 %\sum_{kq} \mathcal{J}_{kq00} 
 %\frac{O_k^q(\hat{\mathbf{J}}_1) \hat{\mathbf{I}}_2}{O_k^0(J_1)}
%&+
 \sum_{kq q'} \mathcal{J}_{kq1q'} 
 \frac{O_k^q({\mathbf{J}}_1) {\rm S}_{2q'}}{O_k^0(J_1)S_2}, 
\label{Eq:HJS}
\end{eqnarray}
where $k$ is odd due to the time reversal symmetry. 

The highest rank $k$ of Stevens operators entering Eqs. (\ref{Eq:HJJ}) and (\ref{Eq:HJS}) is found from the relation \cite{Iwahara2015}:
\begin{eqnarray}
 k_{\rm max} &=& \min\left[2l_1 + 1, 2J_1\right],
\label{Eq:kmax1}
\end{eqnarray}
where $l_1$ is the atomic orbital momentum and 
$J_1$ is the total angular momentum of the ground atomic multiplet on site 1 (a similar relation holds for $k'_{\rm max}$ on site 2). In the case of $f^N$ ions,
$k_{\rm max} = 2l+1 = 7$ for $N=$ 2-4 and 7-13, 
$k_{\rm max} = 2J = 5$ for $N=1,5$, 
and $k_{\rm max} = 0$ for $N=6$.

The maximal component $q$ ($> 0$) in Eqs. (\ref{Eq:HJJ}) and (\ref{Eq:HJS}) is generally determined by details of intersite interaction \cite{Iwahara2015}:
\begin{eqnarray}
 q_{\rm max} &=& \min\left[\Delta_{\rm max}^1+1, k_{\rm max} \right],
\label{Eq:qmax1}
\end{eqnarray}
where $\Delta_{\rm max}^{1}$ is the maximal difference of orbital momentum projections on site 1, $m$ and $n$, entering as indices in the parameters 
$V_{mm'n'n}$ and $t^{12}_{mm'} t^{21}_{n'n}$, defining the direct and the kinetic intersite exchange interaction, respectively ($V$ is the bielectronic integral and $t^{12}$, $t^{21}$ are electron transfer parameters).
For low-symmetric systems, all these parameters will be non-zero, leading to $\Delta_{\rm max}^1 = 2l_1$. Therefore,
in such cases $q_{\rm max}$ is determined by intrasite properties only 
(as $k_{\rm max}$). The maximal value of $q'$ at the second site, $q'_{\rm max}$, is determined in a similar way.

With the above information on the exchange interaction for entire $J$ multiplet,
the effective Hamiltonian is derived.
For more information on the exchange interaction, 
such as the form of $\mathcal{J}_{kqk'q'}$ and explanations of $k_{\rm max}$ and $q_{\rm max}$, 
see Ref. \cite{Iwahara2015}.

\section{Derivation of the effective exchange Hamiltonian} % for axial doublets}
\label{Sec:Exchange_doublets}
In this section, we derive the effective exchange Hamiltonians $\hat{\rm H}_{\rm eff}$ 
between the ground doublets of metal sites and between 
the ground doublet of the metal site and isotropic spin. 
This is done by projecting the complete form of the exchange Hamiltonian for the entire $J$-multiplet, 
Eqs. (\ref{Eq:HJJ}) and (\ref{Eq:HJS}), into the one within the truncated Hilbert space,
and then describe the doublets by the $\tilde{s}=1/2$ pseudospin operators.
We consider first doublets of type (\ref{Eq:axial}), for which the derivation 
%\sout{
%of the exchange Hamiltonian from a complete form of interaction of $J$-multiplets, Eqs. (\ref{Eq:HJJ}) and (\ref{Eq:HJS}), 
%}
is the simplest.

\subsection{Interacting axial doublets}
%We consider first doublets of type (\ref{Eq:axial}), for which the derivation of the exchange Hamiltonian from a complete form of interaction of $J$-multiplets, Eqs. (\ref{Eq:HJJ}) and (\ref{Eq:HJS}), is the simplest. 
Within the truncated Hilbert space, 
$\{|J_1\mu_1 J_2\mu_2\rangle: \mu_i = \pm m_i\}$,
the exchange Hamiltonian (\ref{Eq:HJJ}) reduces to 
\begin{eqnarray}
 \hat{\rm H}_{{\rm eff}} &=& 
 \hat{\rm P}_{12} \hat{\rm H} \hat{\rm P}_{12},
\label{Eq:PHJJP}
\end{eqnarray}
where 
$\hat{\rm P}_{12} = \hat{\rm P}_1 \hat{\rm P}_2$, and 
$\hat{\rm P}_i = \sum_{\mu_i}|J_i\mu_i\rangle \langle J_i\mu_i|$ $(i=1,2)$ is the projector into the truncated space.
The matrix element of the exchange Hamiltonian (\ref{Eq:HJJ}) is 
\begin{eqnarray}
 \langle J_1\mu_1'J_2\mu_2'|\hat{\rm H}|J_1\mu_1J_2\mu_2\rangle 
 &=& 
 \sum_{kk'} \sum_{qq'} \mathcal{J}_{kqk'q'} 
 \frac{C_{J_1\mu_1kq}^{J_1\mu_1'} C_{J_2\mu_2k'q'}^{J_2\mu_2'}}
 {C_{J_1J_1k0}^{J_1J_1} C_{J_2J_2k'0}^{J_2J_2}},
\label{Eq:Hmat}
\end{eqnarray}
%where $M_1',M_1=\pm m_1$, $M_2',M_2 = \pm m_2$. 
using the Wigner-Eckart theorem and $C_{a\alpha b\beta}^{c\gamma}$ is the Clebsch-Gordan coefficients
\cite{Varshalovich1988}
\footnote{We follow the convention of the spherical harmonics and the Clebsch-Gordan coefficients of Ref. \cite{Varshalovich1988}.}.
Thus, substituting Eq. (\ref{Eq:Hmat}) into Eq. (\ref{Eq:PHJJP}), 
the effective Hamiltonian is obtained as 
\begin{eqnarray}
 \hat{\rm H}_{\rm eff}
 &=&
 \sum_{kk'}^{\rm even} \sum_{qq'} 
 \frac{\mathcal{J}_{kqk'q'}}
 {C_{J_1J_1k0}^{J_1J_1} C_{J_2J_2k'0}^{J_2J_2}}
\nonumber\\
 &\times&
 \left[
   C_{J_1m_1k0}^{J_1m_1} 
   \delta_{q,0}
  \left(
    |J_1m_1\rangle \langle J_1m_1|
  + |J_1-m_1\rangle \langle J_1-m_1|
  \right)
\right.
\nonumber\\
 &+&
\left.
   C_{J_1-m_1k2m_1}^{J_1m_1} 
  \left(
    \delta_{q,2m_1}
    |J_1m_1\rangle \langle J_1-m_1|
  +\delta_{q,-2m_1}
    |J_1-m_1\rangle \langle J_1m_1|
  \right)
 \right]
\nonumber\\
 &\times&
 \left[
   C_{J_2m_2k'0}^{J_2m_2} 
   \delta_{q',0}
  \left(
    |J_2m_2\rangle \langle J_2m_2|
  + |J_2-m_2\rangle \langle J_2-m_2|
  \right)
\right.
\nonumber\\
 &+&
\left.
   C_{J_2-m_2k'2m_2}^{J_2m_2} 
  \left(
    \delta_{q',2m_2}
    |J_2m_2\rangle \langle J_2-m_2|
  +\delta_{q',-2m_2}
    |J_2-m_2\rangle \langle J_2m_2|
  \right)
 \right]
\nonumber\\
 &+&
 \sum_{kk'}^{\rm odd} \sum_{qq'} 
 \frac{\mathcal{J}_{kqk'q'}}
 {C_{J_1J_1k0}^{J_1J_1} C_{J_2J_2k'0}^{J_2J_2}}
\nonumber\\
 &\times&
 \left[
   C_{J_1m_1k0}^{J_1m_1} 
   \delta_{q,0}
  \left(
    |J_1m_1\rangle \langle J_1m_1|
   -|J_1-m_1\rangle \langle J_1-m_1|
  \right)
\right.
\nonumber\\
 &+&
\left.
   C_{J_1-m_1k2m_1}^{J_1m_1} 
  \left(
    \delta_{q,2m_1}
    |J_1m_1\rangle \langle J_1-m_1|
  -\delta_{q,-2m_1}
    |J_1-m_1\rangle \langle J_1m_1|
  \right)
 \right]
\nonumber\\
 &\times&
 \left[
   C_{J_2m_2k'0}^{J_2m_2} 
   \delta_{q',0}
  \left(
    |J_2m_2\rangle \langle J_2m_2|
  - |J_2-m_2\rangle \langle J_2-m_2|
  \right)
\right.
\nonumber\\
 &+&
\left.
   C_{J_2-m_2k'2m_2}^{J_2m_2} 
  \left(
    \delta_{q',2m_2}
    |J_2m_2\rangle \langle J_2-m_2|
  -\delta_{q',-2m_2}
    |J_2-m_2\rangle \langle J_2m_2|
  \right)
 \right].
\nonumber\\
\end{eqnarray}
Here, the property of Clebsch-Gordan coefficient related to the time inversion is used \cite{Varshalovich1988}:
\begin{eqnarray}
 C_{a\alpha b\beta}^{c\gamma} &=& (-1)^{a + b - c} C_{a-\alpha b-\beta}^{c-\gamma}.
\label{Eq:CG_1}
\end{eqnarray}
The two doublet functions (\ref{Eq:axial}) on site $i$ can be put in correspondence to the eigenfunctions $|1/2,\pm 1/2\rangle$ of the pseudospin $\tilde{s}_i =1/2$:
%We introduce $1/2$-pseudospin operators for each site:
\begin{eqnarray}
 |J_im_i\rangle \langle J_im_i| &=& \frac{1}{2}\tilde{I}_{i} + \tilde{s}_{i0}
 = \frac{1}{2}\tilde{I}_{i} + \tilde{s}_{iz_i}, 
\\
 |J_i-m_i\rangle \langle J_i-m_i| &=& \frac{1}{2}\tilde{I}_{i} - \tilde{s}_{i0}, 
 = \frac{1}{2}\tilde{I}_{i} - \tilde{s}_{iz_i}, 
\\ 
 |J_im_i\rangle \langle J_i-m_i| &=& -\sqrt{2} \tilde{s}_{i+1} = \tilde{s}_{i+} 
 = \tilde{s}_{ix_i} + i \tilde{s}_{iy_i}, 
\\ 
 |J_i-m_i\rangle \langle J_im_i| &=& \sqrt{2} \tilde{s}_{i-1} = \tilde{s}_{i-} 
 = \tilde{s}_{ix_i} - i \tilde{s}_{iy_i}, 
\end{eqnarray}
where $i=1,2$, $\tilde{I}_{i}$ is the two dimensional unit matrix,
$\tilde{s}_{1\pm}=\tilde{s}_{1x_1} \pm i\tilde{s}_{1y_1}$ and 
$\tilde{s}_{2\pm}=\tilde{s}_{2x_2} \pm i\tilde{s}_{2y_2}$; 
$x_i$ and $y_i$ denote Cartesian axes of site $i$, 
which generally do not coincide with similar axes of another site. 
Using the $1/2$-pseudospin operators,
\begin{eqnarray}
 \hat{\rm H}_{\rm eff}
 &=&
 \sum_{kk'}^{\rm even} \sum_{qq'} 
 \frac{\mathcal{J}_{kqk'q'}}
 {C_{J_1J_1k0}^{J_1J_1} C_{J_2J_2k'0}^{J_2J_2}}
\nonumber\\
 &\times&
 \left[
   C_{J_1m_1k0}^{J_1m_1} 
   \delta_{q,0}
%  \left(
%    |m_1\rangle \langle m_1|
%  + |-m_1\rangle \langle -m_1|
%  \right)
   \tilde{I}_{1}
%\right.
%\nonumber\\
% &+&
%\left.
 +
   C_{J_1-m_1k2m_1}^{J_1m_1} 
  \left(
    \delta_{q,2m_1}
%    |m_1\rangle \langle -m_1|
    \tilde{s}_{1+}
  +\delta_{q,-2m_1}
%    |-m_1\rangle \langle m_1|
    \tilde{s}_{1-}
  \right)
 \right]
\nonumber\\
 &\times&
 \left[
   C_{J_2m_2k'0}^{J_2m_2} 
   \delta_{q',0}
%  \left(
%    |m_2\rangle \langle m_2|
%  + |-m_2\rangle \langle -m_2|
%  \right)
  \tilde{I}_2
%\right.
%\nonumber\\
% &+&
%\left.
 +
   C_{J_2-m_2k'2m_2}^{J_2m_2} 
  \left(
    \delta_{q',2m_2}
%    |m_2\rangle \langle -m_2|
    \tilde{s}_{2+}
  +\delta_{q',-2m_2}
%    |-m_2\rangle \langle m_2|
    \tilde{s}_{2-}
  \right)
 \right]
\nonumber\\
 &+&
 \sum_{kk'}^{\rm odd} \sum_{qq'} 
 \frac{\mathcal{J}_{kqk'q'}}
 {C_{J_1J_1k0}^{J_1J_1} C_{J_2J_2k'0}^{J_2J_2}}
\nonumber\\
 &\times&
 \left[
   C_{J_1m_1k0}^{J_1m_1} 
   \delta_{q,0}
%  \left(
%    |m_1\rangle \langle m_1|
%   -|-m_1\rangle \langle -m_1|
%  \right)
   \frac{\tilde{s}_{1z_1}}{\tilde{s}_1}
%\right.
%\nonumber\\
% &+&
%\left.
 +
   C_{J_1-m_1k2m_1}^{J_1m_1} 
  \left(
    \delta_{q,2m_1}
%    |m_1\rangle \langle -m_1|
    \tilde{s}_{1+}
  -\delta_{q,-2m_1}
%    |-m_1\rangle \langle m_1|
    \tilde{s}_{1-}
  \right)
 \right]
\nonumber\\
 &\times&
 \left[
   C_{J_2m_2k'0}^{J_2m_2} 
   \delta_{q',0}
%  \left(
%    |m_2\rangle \langle m_2|
%  - |-m_2\rangle \langle -m_2|
%  \right)
   \frac{\tilde{s}_{2z_2}}{\tilde{s}_2}
%\right.
%\nonumber\\
% &+&
%\left.
 +
   C_{J_2-m_2k'2m_2}^{J_2m_2} 
  \left(
    \delta_{q',2m_2}
%    |m_2\rangle \langle -m_2|
    \tilde{s}_{2+}
  -\delta_{q',-2m_2}
%    |-m_2\rangle \langle m_2|
    \tilde{s}_{2-}
  \right)
 \right].
\label{Eq:Heff_spin}
\end{eqnarray}
%Finally, the effective Hamiltonian for two doublets is obtained of the form:
%%
%the matrix of 
Finally, the general exchange interaction (\ref{Eq:HJJ}) projected onto doublet functions of two sites 
can be recast in the following pseudospin exchange Hamiltonian:
\begin{eqnarray}
 \hat{\rm H}_{\rm eff}
 &=&
   \mathcal{J}_{II} \tilde{I}_1 \tilde{I}_2 
% + \mathcal{K}_{zz} \frac{\tilde{s}_{1z_1}}{\tilde{s}_1} \frac{\tilde{s}_{2z_2}}{\tilde{s}_2}
 + 4\mathcal{K}_{zz} \tilde{s}_{1z_1}\tilde{s}_{2z_2}
\nonumber\\
 &+&
   \mathcal{J}_{I+} \tilde{I}_1 \tilde{s}_{2+} 
 + \mathcal{J}_{I-} \tilde{I}_1 \tilde{s}_{2-} 
 + \mathcal{J}_{+I} \tilde{s}_{1+} \tilde{I}_{2} 
 + \mathcal{J}_{-I} \tilde{s}_{1-} \tilde{I}_{2} 
\nonumber\\
 &+&
   2\mathcal{K}_{z+} \tilde{s}_{1z_1} \tilde{s}_{2+} 
 + 2\mathcal{K}_{z-} \tilde{s}_{1z_1} \tilde{s}_{2-} 
 + 2\mathcal{K}_{+z} \tilde{s}_{1+} \tilde{s}_{2z_2}
 + 2\mathcal{K}_{-z} \tilde{s}_{1-} \tilde{s}_{2z_2}
\nonumber\\
 &+&
   \left(\mathcal{J}_{++} + \mathcal{K}_{++}\right) \tilde{s}_{1+} \tilde{s}_{2+} 
 + \left(\mathcal{J}_{+-} + \mathcal{K}_{+-}\right) \tilde{s}_{1+} \tilde{s}_{2-} 
\nonumber\\
 &+&
   \left(\mathcal{J}_{-+} + \mathcal{K}_{-+}\right) \tilde{s}_{1-} \tilde{s}_{2+} 
 + \left(\mathcal{J}_{--} + \mathcal{K}_{--}\right) \tilde{s}_{1-} \tilde{s}_{2-}.
\label{Eq:H_Lines_2}
\end{eqnarray}
%
%\sout{
%Here, $\mathcal{J}$ and $\mathcal{K}$ are constants. 
%The effective Hamiltonian (\ref{Eq:H_Lines_2}) contains the constant part (the first term), 
%the zero-field splitting (ZFS) part (terms 3-6) and the exchange part (the rest of the terms). 
%In Eq. (\ref{Eq:H_Lines_2}),} 
Here, the parameters $\mathcal{J}$ and $\mathcal{K}$ 
are combinations of exchange parameters $\mathcal{J}_{kqk'q'}$ from Eq. (\ref{Eq:HJJ}) 
and products of Clebsch-Gordan coefficients: % (see \ref{A:JK}).
\begin{eqnarray}
 \mathcal{J}_{II} &=& 
 \sum_{kk'}^{\rm even} 
 \mathcal{J}_{k0k'0}
 \frac{C_{J_1m_1k0}^{J_1m_1}C_{J_2m_2k'0}^{J_2m_2}}{C_{J_1J_1k0}^{J_1J_1} C_{J_2J_2k'0}^{J_2J_2}},
 \label{Eq:JII}
\\
 \mathcal{J}_{I\pm} &=& 
% (-1)^{2m_2} 
 \mathcal{J}_{I\mp}^*
% = (-1)^{2m_2} \mathcal{J}_{I\pm}
%\nonumber\\
% &=&
 =
  \frac{1+(-1)^{2m_2}}{2}
  \sum_{kk'}^{\rm even} 
 \mathcal{J}_{k0k'\pm 2m_2}
 \frac{C_{J_1m_1k0}^{J_1m_1}C_{J_2-m_2k'2m_2}^{J_2m_2}}
 {C_{J_1J_1k0}^{J_1J_1} C_{J_2J_2k'0}^{J_2J_2}},
%\nonumber\\
% &=& (-1)^{2m_2} \left(\mathcal{J}_{I\mp}\right)^*
% = (-1)^{2m_2} \mathcal{J}_{I\pm}
 \label{Eq:JIpm}
\\
 \mathcal{J}_{\pm I} 
 &=& %(-1)^{2m_1} 
 \mathcal{J}_{\mp I}^*
% &=& 
 = \frac{1+(-1)^{2m_1}}{2}
 \sum_{kk'}^{\rm even} 
 \mathcal{J}_{k\pm 2m_1k'0}
 \frac{C_{J_1-m_1k2m_1}^{J_1m_1} C_{J_2m_2k'0}^{J_2m_2}} 
  {C_{J_1J_1k0}^{J_1J_1} C_{J_2J_2k'0}^{J_2J_2}},
%\nonumber\\
% &=& (-1)^{2m_1} \left(\mathcal{J}_{\mp I}\right)^*
% = (-1)^{2m_1} \mathcal{J}_{\pm I}
 \label{Eq:JpmI}
\\
 \mathcal{J}_{\pm \pm} 
 &=&
 %(-1)^{2m_1+2m_2} 
 \mathcal{J}_{\mp \mp}^*
\nonumber\\
 &=& \frac{1+(-1)^{2(m_1+m_2)}}{2}
 \sum_{kk'}^{\rm even} 
 \mathcal{J}_{k\pm 2m_1k'\pm 2m_2}
 \frac{C_{J_1-m_1k2m_1}^{J_1m_1} C_{J_2-m_2k'2m_2}^{J_2m_2}} 
  {C_{J_1J_1k0}^{J_1J_1} C_{J_2J_2k'0}^{J_2J_2}},
%\nonumber\\
% &=& (-1)^{2m_1+2m_2} \left(\mathcal{J}_{\mp \mp}\right)^*
% = (-1)^{2m_1+2m_2} \mathcal{J}_{\pm \pm}
% \sum_{kk'}^{\rm even} 
% \frac{1+(-1)^{k+2m_1}}{2}
% \frac{1+(-1)^{k'+2m_2}}{2}
% \mathcal{J}_{k\pm 2m_1k'\pm 2m_2}
% \frac{C_{J_1-m_1k2m_1}^{J_1m_1} C_{J_2-m_2k'2m_2}^{J_2m_2}} 
%  {C_{J_1J_1k0}^{J_1J_1} C_{J_2J_2k'0}^{J_2J_2}},
 \label{Eq:Jpmpm}
\\
 \mathcal{J}_{\pm \mp} &=& 
 %(-1)^{2m_1+2m_2} 
 \mathcal{J}_{\mp \pm}^*
\nonumber\\
 &=& 
 \frac{1+(-1)^{2(m_1+m_2)}}{2}
 \sum_{kk'}^{\rm even} 
 \mathcal{J}_{k\pm 2m_1k'\mp 2m_2}
 \frac{C_{J_1-m_1k2m_1}^{J_1m_1} C_{J_2-m_2k'2m_2}^{J_2m_2}} 
  {C_{J_1J_1k0}^{J_1J_1} C_{J_2J_2k'0}^{J_2J_2}},
%\nonumber\\
% &=& (-1)^{2m_1+2m_2} \left(\mathcal{J}_{\mp \pm}\right)^*
% = (-1)^{2m_1+2m_2} \mathcal{J}_{\pm \mp},
%&=&
% \sum_{kk'}^{\rm even} 
% \frac{1+(-1)^{k+2m_1}}{2}
% \frac{1+(-1)^{k'+2m_2}}{2}
% \mathcal{J}_{k\pm 2m_1k'\mp 2m_2}
% \frac{C_{J_1-m_1k2m_1}^{J_1m_1} C_{J_2-m_2k'2m_2}^{J_2m_2}} 
%  {C_{J_1J_1k0}^{J_1J_1} C_{J_2J_2k'0}^{J_2J_2}}
 \label{Eq:Jpmmp}
\\
 \mathcal{K}_{zz} &=& 
 \sum_{kk'}^{\rm odd} \mathcal{J}_{k0k'0}
   \frac{C_{J_1m_1k0}^{J_1m_1} C_{J_2m_2k'0}^{J_2m_2}}
   {C_{J_1J_1k0}^{J_1J_1} C_{J_2J_2k'0}^{J_2J_2}},
 \label{Eq:Kzz}
\\
 \mathcal{K}_{z\pm} &=& 
%  -(-1)^{2m_2} 
 \mathcal{K}_{z\mp}^*
 = 
 \pm \frac{1-(-1)^{2m_2}}{2}
 \sum_{kk'}^{\rm odd} \mathcal{J}_{k0k'\pm 2m_2}
   \frac{C_{J_1m_1k0}^{J_1m_1} C_{J_2-m_2k'2m_2}^{J_2m_2}}
   {C_{J_1J_1k0}^{J_1J_1} C_{J_2J_2k'0}^{J_2J_2}},
%\nonumber\\
% &=&
%   -(-1)^{2m_2} \left(\mathcal{K}_{z\mp}\right)^*
%  = -(-1)^{2m_2} \mathcal{K}_{z\pm},
 \label{Eq:Kzpm}
\\
 \mathcal{K}_{\pm z} &=& 
  \mathcal{K}_{\mp z}^*
 = \pm \frac{1-(-1)^{2m_1}}{2}
 \sum_{kk'}^{\rm odd} \mathcal{J}_{k\pm 2m_1k'0}
   \frac{C_{J_1-m_1k2m_1}^{J_1m_1} C_{J_2m_2k'0}^{J_2m_2}}
   {C_{J_1J_1k0}^{J_1J_1} C_{J_2J_2k'0}^{J_2J_2}},
%\nonumber\\
% &=&
%    -(-1)^{2m_1} \left(\mathcal{K}_{\mp z}\right)^*
%  = -(-1)^{2m_1} \mathcal{K}_{\pm z},
 \label{Eq:Kpmz}
\\
 \mathcal{K}_{\pm \pm} &=& 
  \mathcal{K}_{\mp \mp}^* 
\nonumber\\
 &=&
 \frac{1+(-1)^{2(m_1+m_2)}}{2}
 \sum_{kk'}^{\rm odd} \mathcal{J}_{k\pm 2m_1k'\pm 2m_2}
   \frac{C_{J_1-m_1k2m_1}^{J_1m_1} C_{J_2-m_2k'2m_2}^{J_2m_2}}
   {C_{J_1J_1k0}^{J_1J_1} C_{J_2J_2k'0}^{J_2J_2}},
%\nonumber\\
% &=&
%   (-1)^{2m_1+2m_2} \left(\mathcal{K}_{\mp \mp}\right)^* 
% = (-1)^{2m_1+2m_2} \mathcal{K}_{\pm \pm},
 \label{Eq:Kpmpm}
\\
 \mathcal{K}_{\pm \mp} &=& 
 \mathcal{K}_{\mp \pm}^*
\nonumber\\
 &=&
 - \frac{1+(-1)^{2(m_1+m_2)}}{2} \sum_{kk'}^{\rm odd} \mathcal{J}_{k\pm 2m_1k'\mp 2m_2}
   \frac{C_{J_1-m_1k2m_1}^{J_1m_1} C_{J_2-m_2k'2m_2}^{J_2m_2}}
   {C_{J_1J_1k0}^{J_1J_1} C_{J_2J_2k'0}^{J_2J_2}}.
%\nonumber\\
% &=&
%   (-1)^{2m_1+2m_2} \left(\mathcal{K}_{\mp \pm}\right)^*
% = (-1)^{2m_1+2m_2} \mathcal{K}_{\pm \mp}.
 \label{Eq:Kpmmp}
\end{eqnarray}
The above relations between exchange parameters are derived using 
%The relations in Eqs.  (\ref{Eq:JIpm}), (\ref{Eq:JpmI}), (\ref{Eq:Jpmpm}), (\ref{Eq:Jpmmp}), 
%(\ref{Eq:Kzpm}), (\ref{Eq:Kpmz}), (\ref{Eq:Kpmpm}), (\ref{Eq:Kpmmp})
%are derived in \ref{A:Hmat}.
%For the derivation, time reversal symmetry is used.
%$\mathcal{J}$ and $\mathcal{K}$ contain only $\mathcal{J}_{kqk'q'}$ with even and odd $k$, $k'$, respectively.
%Using the relation 
%\sout{
%the property $\left(\mathcal{J}_{kqk'q'}\right)^* = (-1)^{q + q'} \mathcal{J}_{k-qk'-q'}$ \cite{Iwahara2015}
%}
the time reversal symmetry \cite{Abragam1970}
and the symmetry relations for Clebsch-Gordan coefficients \cite{Varshalovich1988} (see \ref{A:Hmat}).

The effective Hamiltonian (\ref{Eq:H_Lines_2}) contains the constant part (the first term), 
the zero-field splitting (ZFS) part (terms 3-6) and the exchange part (the rest of the terms). 
With above equations we are able to identify the situations when $\mathcal{J}$ and $\mathcal{K}$ become zero
%\sout{
%The ZFS terms only appear when at least one metal ion is of non-Kramers type (integer $J_1$ and $J_2$) and look artifactual (are not invariant under time reversal) because we describe non-Kramers ions by pseudospins $\tilde{s}=1/2$. 
%This drawback disappear when the description of such ions is done with integer pseudospins, e.g., $\tilde{s}=1$.
%This is not necessary here because the exchange part (which is only of interest here) is described correctly already within such a simplified treatment.
%}
{
\footnote{
The ZFS terms only appear when at least one metal ion is of non-Kramers type (integer $J_1$ and $J_2$) and look artifactual (are not invariant under time reversal) because we describe non-Kramers ions by pseudospins $\tilde{s}=1/2$. 
This drawback disappear when the description of such ions is done with integer pseudospins, e.g., $\tilde{s}=1$ \cite{Mueller1968}. 
This is not necessary here because the exchange part (which is only of interest here) is described correctly already within such a simplified treatment.
}.
The overview of such situations in function of the parity of electrons number on sites ($N_1$ and $N_2$) 
and of the relation between $2m_i$ ($i=1,2$) and $q_{\rm max}, q'_{\rm max}$ (\ref{Eq:qmax1}) is given in \ref{A:JK}.
This information is sufficient to find general conditions under which the exchange Hamiltonian for two interacting doublets (\ref{Eq:H_Lines_2}) becomes of Ising type.

\subsection{Interacting axial doublet and isotropic spin}
In the case of a doublet (\ref{Eq:axial}) interacting with an isotropic spin, the derivations are similar.
%\sout{
%, except that now} 
Now,
we only need to consider the matrix elements of operators referring to the first site in Eq. (\ref{Eq:HJS}):
%Starting from general exchange interaction between a $J$-multiplet and an isotropic spin, 
%
%\begin{eqnarray}
% \hat{\rm H} &=& \sum_{k}^{\rm odd} \sum_{q=-k}^k \sum_{q'=-1}^1 
% \mathcal{J}_{kq1q'} \frac{O_k^q({\mathbf{J}}_1){\rm S}_{2q'}}{O_k^0(J_1)S_2}.
%\end{eqnarray}
%
%the Hilbert space for site 1 is downfolded onto the doublet states ($m_1>0$): 
%\begin{eqnarray}
% \{|J_1-m_1\rangle, |J_1m_1\rangle \}.
%\end{eqnarray}
\begin{eqnarray}
 \hat{{\rm H}}_{\rm eff} &=& \hat{\rm P}_1 \hat{\rm H} \hat{\rm P}_1.
% \hat{{\rm H}}_{\rm eff} &=& \sum_{\mu_1,\mu'_1=\pm m_1} 
% |J_1 \mu_1'\rangle \langle J_1 \mu_1'|\hat{\rm H}|J_1 \mu_1\rangle \langle J_1 \mu_1|.
\end{eqnarray}
The matrix elements of $\hat{\rm H}$ (\ref{Eq:HJS}) are given as 
\begin{eqnarray}
 \langle J_1\mu_1'|\hat{\rm H}|J_1\mu_1\rangle &=& 
 \sum_{kqq'} \mathcal{J}_{kq1q'} \frac{C_{J_1\mu_1kq}^{J_1\mu_1'} {\rm S}_{2q'}}
 {C_{J_1J_1k0}^{J_1J_1}S_2},
\end{eqnarray}
%Therefore, the effective Hamiltonian is 
%\begin{eqnarray}
% \hat{\rm H}_{\rm eff} &=& \sum_{k}^{\rm odd}\sum_{q'=-1}^1 
% \left[
%  \mathcal{J}_{k01q'} \frac{C_{J_1m_1k0}^{J_1m_1}}{C_{J_1J_1k0}^{J_1J_1}} 
% \left(|J_1m_1\rangle \langle J_1m_1| - |J_1-m_1\rangle \langle J_1-m_1|\right)
% \right.
%\nonumber\\
%&+&
% \mathcal{J}_{k2m_11q'} \frac{C_{J_1-m_1k2m_1}^{J_1m_1}}{C_{J_1J_1k0}^{J_1J_1}} 
% |J_1m_1\rangle \langle J_1-m_1|
%\nonumber\\
%&-& 
% \left.
% \mathcal{J}_{k-2m_11q'} \frac{C_{J_1-m_1k2m_1}^{J_1m_1}}{C_{J_1J_1k0}^{J_1J_1}} 
% |J_1-m_1\rangle \langle J_1m_1|
% \right]
% \frac{{\rm S}_{2q'}}{S_2}
%\nonumber\\
% &=&
% \sum_{q'=-1}^1 
% \left[
%  \mathcal{K}_{zq'} \frac{\tilde{s}_{1z_1}}{\tilde{s}_1}
%+ \mathcal{K}_{+q'} \tilde{s}_{1+}
%+ \mathcal{K}_{-q'} \tilde{s}_{1-}
% \right]
% \frac{{\rm S}_{2q'}}{S_2},
%\end{eqnarray}
%where 
%\begin{eqnarray}
% \mathcal{K}_{zq'} &=& \sum_k^{\rm odd} \mathcal{J}_{k01q'} 
% \frac{C_{J_1m_1k0}^{J_1m_1}}{C_{J_1J_1k0}^{J_1J_1}}, 
%\\
% \mathcal{K}_{\pm q'} &=&
% \pm \sum_{k}^{\rm odd} \mathcal{J}_{k\pm 2m_11q'} \frac{C_{J_1-m_1k2m_1}^{J_1m_1}}{C_{J_1J_1k0}^{J_1J_1}}.
%\end{eqnarray}
and the projected exchange Hamiltonian looks as follows: % \cite{SM}:
\begin{eqnarray}
 \hat{\rm H}_{\rm eff} &=&
\sum_{q'=-1}^1 
 \left[
  2\mathcal{K}_{zq'} \tilde{s}_{1z_1}
+ \mathcal{K}_{+q'} \tilde{s}_{1+}
+ \mathcal{K}_{-q'} \tilde{s}_{1-}
 \right]
 \frac{{\rm S}_{2q'}}{S_2}. 
\label{Eq:H_doub_iso}
\end{eqnarray}
%
%\sout{
%where} 
Here,
spherical components for ${\mathbf{S}}_{2}$ are used \cite{Varshalovich1988} 
and the expressions for parameters $\mathcal{K}$ are given as %in \ref{A:JK}. 
\begin{eqnarray}
 \mathcal{K}_{zq'} &=& \sum_k^{\rm odd} \mathcal{J}_{k01q'} 
 \frac{C_{J_1m_1k0}^{J_1m_1}}{C_{J_1J_1k0}^{J_1J_1}}, 
\\
 \mathcal{K}_{\pm q'} &=&
 \pm \sum_{k}^{\rm odd} \mathcal{J}_{k\pm 2m_11q'} \frac{C_{J_1-m_1k2m_1}^{J_1m_1}}{C_{J_1J_1k0}^{J_1J_1}}.
\end{eqnarray}
As in the previous case, Eq. (\ref{Eq:H_Lines_2}), the exchange Hamiltonian (\ref{Eq:H_doub_iso}) 
can become of Ising type under certain conditions.

\section{Two types of Ising exchange interaction}
\label{Sec:type_Ising}
With the use of the properties of $\mathcal{J}$ and $\mathcal{K}$ in Sec. \ref{Sec:Exchange_doublets},
we discuss the form of the effective pseudospin Hamiltonian.
Based on the analysis, we show that there are two types of the Ising Hamiltonian.

Consider as example a situation when both $N_1$ and $N_2$ are odd. For $m_1 > q_{\rm max}/2$ and $m_2 > q'_{\rm max}/2$, we obtain all parameters in Eq. (\ref{Eq:H_Lines_2}) zero except $\mathcal{K}_{zz}$, resulting in
\begin{eqnarray}
 \hat{\rm H}_{\rm I} &=&
  % \mathcal{J}_{II} \tilde{I}_1 \tilde{I}_2 
 4\mathcal{K}_{zz} \tilde{s}_{1z_1} \tilde{s}_{2z_2},
\label{Eq:H_Lines_21}
\end{eqnarray}
i.e. in an Ising Hamiltonian of coaxial type (\ref{Eq:H_coax}), with an ordering of magnetic moments shown in the left plot of Fig. \ref{Fig1}(a). 

If we diminish the axiality of site 2 %$m_2$ 
so that $m_1 > q_{\rm max}/2$ and $m_2 \le q'_{\rm max}/2$, 
Eq. (\ref{Eq:H_Lines_2}) reduces to
\begin{eqnarray}
 \hat{\rm H}_{\rm eff} &=&
 4 \mathcal{K}_{zz} \tilde{s}_{1z_1}\tilde{s}_{2z_2}
 + 2\mathcal{K}_{z+} \tilde{s}_{1z_1} \tilde{s}_{2+} 
 + 2\mathcal{K}_{z-} \tilde{s}_{1z_1} \tilde{s}_{2-},\;\;\;\;\;
\label{Eq:H_Lines_22}
\end{eqnarray}
where all terms contain $\tilde{s}_{1z_1}$. The expression multiplying the latter is a combination of pseudospin operators of the second site which can be written in the form
\begin{eqnarray}
\tilde{s}_{2z'} &=& 
   \frac{\Re(\mathcal{K}_{z+})}{\mathcal{K}'_{zz'}}\tilde{s}_{2x_2} 
 - \frac{\Im(\mathcal{K}_{z+})}{\mathcal{K}'_{zz'}}\tilde{s}_{2y_2}
 + \frac{\mathcal{K}_{zz}}{\mathcal{K}'_{zz'}}\tilde{s}_{2z_2},  
\nonumber\\
\mathcal{K}'_{zz'} &=& \sqrt{\mathcal{K}_{zz}^2+\left|\mathcal{K}_{z+}\right|^2}.
\label{Eq:pseudo_2}
\end{eqnarray}
Since the coefficients in the first equation are real and normalized to unity, the corresponding combination of pseudospin projections can be viewed as a rotated pseudospin from initial direction $z_2$ towards a new direction $z'$. Accordingly, the coefficients are directional cosines of $z'$ axis in the coordinate system of site 2. Despite the fact that pseudospin is not related to any physical angular momentum but only defined via Pauli matrices in the basis of two states (\ref{Eq:axial}), one can still define its rotation via induced transformations of doublet wave functions \cite{Chibotaru2013}. Then the eigenfunctions of the new pseudospin operator $\tilde{s}_{2z'}$ will be the same functions (\ref{Eq:axial}) defined with respect to rotated quantization axis $z'$, i.e., $|J_i \pm m_i \rangle'$. Then Eq. (\ref{Eq:H_Lines_22}) is rewritten as 
\begin{eqnarray}
 \hat{\rm H}_{\rm II} &=&
4\mathcal{K}'_{zz'} \tilde{s}_{1z_1} \tilde{s}_{2z'},
\label{Eq:H_II}
\end{eqnarray}
which has the form of an Ising Hamiltonian of {\em non-coaxial} type because the magnetization at the second site will not be directed along the corresponding main magnetic axis $z_2$ but along a different axis $z'$ (left plot in Fig. \ref{Fig1}(b)). 
This makes it qualitatively different from the coaxial Ising interaction described by Eqs. (\ref{Eq:H_coax}), (\ref{Eq:H_Lines_21}). 
Accordingly, the two Ising interactions will be called hereafter of type I and II. 
%The other cases where at least one of $N_1$ and $N_2$ is even are described in the Supplemental Materials \cite{SM}.

On the other hand, the interaction of a doublet (\ref{Eq:axial}) with an isotropic spin, Eq. (\ref{Eq:H_doub_iso}), can only become of Ising type if both  parameters $\mathcal{K}_{+q'}$ and $\mathcal{K}_{-q'}$ are zero. This is achieved when $m_1 > q_{\rm max}/2$, in which case the Hamiltonian (\ref{Eq:H_doub_iso}) becomes of the form:
\begin{eqnarray}
\hat{\rm H}_{\rm II} &=&
   2\mathcal{K}'_{zz'} 
   \tilde{s}_{1z_1}
   \frac{{\rm S}_{2z'}}{S_2}, 
\nonumber\\
%\mathcal{K}'_{zz'} &=& \sqrt{\sum_{q'=-1}^1 |\mathcal{K}_{zq'}|^2},
\mathcal{K}'_{zz'} &=& \sqrt{|\mathcal{K}_{z0}|^2+|\mathcal{K}_{z+1}|^2+|\mathcal{K}_{z-1}|^2},
\label{Eq:H_II_pseudo_S}
\end{eqnarray} 
in which the combination
\begin{eqnarray}
{\rm S}_{2z'} &=& \sum_{q'=-1}^1\frac{\mathcal{K}_{zq'} {\rm S}_{2q'}}{\mathcal{K}'_{zz'}}
 \label{Eq:rotated_S} 
\end{eqnarray}	
is a rotated spin ${\rm S}_{2z}$ from direction $z_1$ to $z'$. 
This means that the Hamiltonian (\ref{Eq:H_II_pseudo_S}) is of non-coaxial Ising type, in which the magnetic moment on site 2 points into a direction different from the main magnetic axis on site 1, as indicated in the right plot of Fig. \ref{Fig1}(b) (cf. Fig. \ref{Fig1}(a)). 
Note that the Ising interaction of type I is only achieved when %the parameters $\propto$
$\mathcal{K}_{z\pm 1}$ in Eq. (\ref{Eq:rotated_S}) are both zero, which requires special symmetry of the exchange bridge.  
%$\mathcal{K}_{z+1}$ and $\mathcal{K}_{z-1}$ in (\ref{Eq:rotated_S}) are both zero, which requires special symmetry of the exchange bridge.  

%\begin{table}
%\caption{\label{label}Table caption.}
%\begin{indented}
%\item[]\begin{tabular}{@{}llll}
%\br
%Head 1&Head 2&Head 3&Head 4\\
%\mr
%1.1&1.2&1.3&1.4\\
%2.1&2.2&2.3&2.4\\
%\br
%\end{tabular}
%\end{indented}
%\end{table}

\begin{table}[tbh]
\caption{
\label{Table1}
Exchange interaction between two doublets and between a doublet and an isotropic spin. 
%{\footnote {I and II denote the two types of Ising exchange interaction, Eqs. (\ref{Eq:H_coax}, \ref{Eq:H_Lines_21}) 
%and (\ref{Eq:H_II}, \ref{Eq:H_II_pseudo_S}), respectively; N stands for non-Ising exchange interaction.}} 
I and II denote the two types of Ising exchange interaction, Eqs. (\ref{Eq:H_coax}), (\ref{Eq:H_Lines_21})
and Eqs. (\ref{Eq:H_II}), (\ref{Eq:H_II_pseudo_S}), respectively; N stands for non-Ising exchange interaction.
}
\begin{indented}
\item[]\begin{tabular}{@{}cccccc}
\br
\multirow{2}{*}{\backslashbox{$N_2$}{$N_1$}}
&& \multicolumn{2}{c}{odd} & \multicolumn{2}{c}{even} \\
&& $m_1 > q_{\rm max}/2$ & $m_1 \le q_{\rm max}/2$ & $m_1> q_{\rm max}/2$ & $m_1 \le q_{\rm max}/2$ \\
\mr
odd  & $m_2 >   q'_{\rm max}/2$   & I  & II & I  & N \\
odd  & $m_2 \le q'_{\rm max}/2$   & II & N  & II & N \\
even & $m_2 >   q'_{\rm max}/2$   & I  & II & I  & N \\
even & $m_2 \le q'_{\rm max}/2$   & N  & N  & N  & N \\
\multicolumn{2}{c}{isotropic spin} & II & N  & II & N \\
\br
\end{tabular}
\end{indented}
\end{table}

The classification for all cases are shown in \ref{A:type_Ising} and 
Table \ref{Table1} summarizes our main result. It shows that rule 2) (Ising interaction of type I) is not satisfied in most cases.
We can see also that the type of resulting exchange Hamiltonian merely depends on the relative values of $m_1$, $m_2$ and $q_{\rm max}/2$, $q'_{\rm max}/2$, respectively. 
Then we can  generalize the results in Table \ref{Table1} over arbitrary doublets (\ref{Eq:doublet}) if by $m_1$ and $m_2$ we will understand not particular doublets (\ref{Eq:axial}) but the minimal absolute values of the index $m$ in the expansion of the corresponding doublet wave function $|M\rangle$ in Eq. (\ref{Eq:doublet}).

\begin{figure}[tb] 
\begin{center}
\includegraphics[height=6cm, bb=0 0 1221 1083]{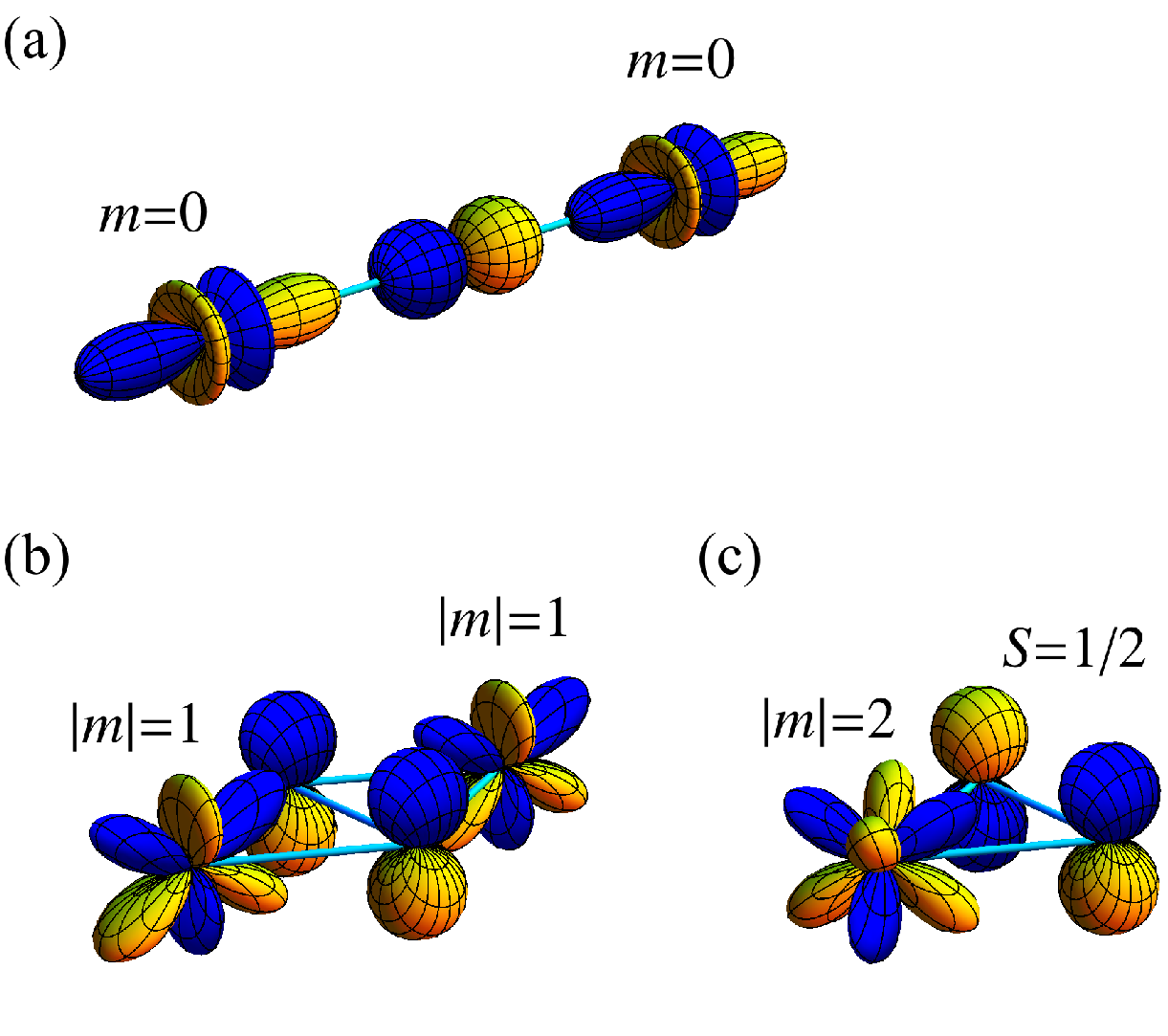}
\end{center}
\caption{(Color online) Examples of symmetric superexchange interaction: (a) via $p_{\sigma}$ orbital of a single bridging atom, connecting only $m=0$ orbitals on lanthanide sites; (b) via the highest occupied orbital of the N$_2^{2-}$ bridge \cite{Rinehart2011}, promoting interaction between $m=\pm 1$ orbitals on lanthanide sites; (c) unpaired electron ($S=1/2$) at N$_2^{3-}$ \cite{Rinehart2011} interacting only with $m=\pm 2$ orbitals of lanthanide due to symmetry restrictions.}
\label{Fig2}
\end{figure}

\section{Ising metal ions}
\label{Sec:Ising_ions}
Table \ref{Table1} shows that rule 1) does not hold either. 
Instead, the realization of Ising exchange interaction depends on the value of $q_{\rm max}$ ($q'_{\rm max}$) 
which is determined by the smaller of the quantities $\Delta_{\rm max} +1$ and $k_{\rm max}$ (Eq. (\ref{Eq:qmax1})). 
$\Delta_{\rm max}$ is small only for sufficiently high symmetry of the exchange bridge. 
For instance, in the case of linear bridge (Fig. \ref{Fig2}(a)) it can have the smallest possible value, $\Delta_{\rm max}=0$. 
Its value increases fast with lowering of the symmetry of the bridge, and the accompanying increase of the mixing of $m$ orbitals, 
being 2 for the case shown in Fig. \ref{Fig2}(b) and 4 for the case shown in Fig. \ref{Fig2}(c). 
When $\Delta_{\rm max}^{1,2}$ on sites become sufficiently large, $q_{\rm max}$ ($q'_{\rm max}$) is determined by 
$k_{\rm max}$ ($k'_{\rm max}$), Eq. (\ref{Eq:kmax1}), i.e., the smaller of the quantities $2l_i + 1$ ($=7$ for $f$ orbitals) 
and $2J_i$ on corresponding sites. 
Designing a strongly axial crystal field on sites one can achieve $m_i$ values as high as $J_i$ \cite{Chibotaru2015}. 
Then such ions with $J_i >7/2$  will {\it a priori} provide Ising exchange interaction with any other magnetic site according to Table \ref{Table1}. 
Contrary to what is stated by the rule 1), these are precisely the ones which can be called Ising ions. They are listed in Table \ref{Table2}.

\begin{table}[tbh]
\caption{
\label{Table2}
Ising metal ions. 
%\footnote{ Only valencies of metal ions in known complexes are listed \cite{Wybourne1965, Edelstein2006}.}
Only valencies of metal ions in known complexes are listed \cite{Wybourne1965, Edelstein2006}.
}
\begin{indented}
\item[]\begin{tabular}{@{}ccccccccc}
\br
\backslashbox{$M^{n+}$}{$f^N$} & $f^2$     & $f^3$     & $f^4$     & $f^8$     & $f^9$     & $f^{10}$  & $f^{11}$  & $f^{12}$ \\
\mr
 Ln$^{3+}$                     & Pr$^{3+}$ & Nd$^{3+}$ & Pm$^{3+}$ & Tb$^{3+}$ & Dy$^{3+}$ & Ho$^{3+}$ & Er$^{3+}$ & Tm$^{3+}$ \\
 Ac$^{2+}$                     &           &           &           &           &           &           & Es$^{2+}$ &            \\
 Ac$^{3+}$                     &           & U$^{3+}$  & Np$^{3+}$ & Bk$^{3+}$ & Cf$^{3+}$ & Es$^{3+}$ &           &           \\
 Ac$^{4+}$                     & U$^{4+}$  & Np$^{4+}$ & Pu$^{4+}$ & Cf$^{4+}$ &           &           &           &           \\
 Ac$^{5+}$                     & Np$^{5+}$ & Pu$^{5+}$ &           &           &           &           &           &           \\
 Ac$^{6+}$                     & Pu$^{6+}$ &           &           &           &           &           &           &           \\
\br
\end{tabular}
\end{indented}
\end{table}

\section{Conclusions}
\label{Sec:Conclusions}
In this work, we find necessary and sufficient conditions to achieve Ising exchange interaction in materials involving lanthanide and actinide metal ions. Unless the symmetry of the exchange bridge is not very high, these conditions are determined solely by the electronic properties of individual metal ions. According to these conditions, by far not all metal ions can display Ising exchange interaction (Table \ref{Table2}). We also find that two types of Ising exchange interaction can arise in these systems, the coaxial and non-coaxial ones. In particular, it is established that the Ising interaction between an anisotropic doublet and an isotropic spin is of second type. 
The basic property of Ising coupling is the lack of dynamics in resulting exchange doublets, also in applied magnetic field. This condition is indispensable for achieving single-molecule magnets since it ensures the quenching of quantum tunneling of magnetization in individual exchange doublets.
The insight gained in this work will contribute to purposeful design of lanthanide and actinide based materials.

{\it Note added}. The Ising exchange interaction between strongly axial ions (Dy, Ho) has been
recently addressed in the preprint \cite{Rau2015} within a different approach from the present work. 
%The main conclusion agrees with the present work.

\appendix
\section{Exchange parameters for the effective model}
The relations between exchange parameters, 
(\ref{Eq:JII})-(\ref{Eq:Kpmmp}),
%(\ref{Eq:JIpm}),
%(\ref{Eq:JpmI}),
%(\ref{Eq:Jpmpm}),
%(\ref{Eq:Jpmmp}),
%(\ref{Eq:Kzz}),
%(\ref{Eq:Kzpm}),
%(\ref{Eq:Kpmz}),
%(\ref{Eq:Kpmpm}),
%(\ref{Eq:Kpmmp}),
are derived here.
In the derivation, we use the property of the exchange parameter under time inversion \cite{Iwahara2015},
\begin{eqnarray}
 \mathcal{J}_{kqk'q'}^* = (-1)^{q + q'} \mathcal{J}_{k-qk'-q'},
\label{Eq:Jcc}
\end{eqnarray}
and the property of the Clebsch-Gordan coefficients \cite{Varshalovich1988},
\begin{eqnarray}
 C_{a\alpha b\beta}^{c\gamma} &=& (-1)^{b+\beta} \sqrt{\frac{2c+1}{2a+1}} C_{c-\gamma b\beta}^{a-\alpha}.
\label{Eq:CG_2}
\end{eqnarray}

\subsection{Exchange parameters of $\hat{\rm H}_{\rm eff}$}
\label{A:Hmat}
The relations between the exchange parameters of $\hat{\rm H}_{\rm eff}$,
$\mathcal{J}$ and $\mathcal{K}$ (\ref{Eq:JII})-(\ref{Eq:Kpmmp}), are derived.
We obtain the form of $\mathcal{J}$ and $\mathcal{K}$ expanding Eq. (\ref{Eq:Heff_spin}).
Among the exchange parameters, $\mathcal{J}_{II}$ (\ref{Eq:JII}) and $\mathcal{K}_{zz}$ (\ref{Eq:Kzz}) 
are obtained by this direct calculation, 
whereas for the others, additional calculations with Eqs. (\ref{Eq:Jcc}) and (\ref{Eq:CG_2}) are required.
\begin{itemize}
 \item $\mathcal{J}_{I\pm}$ and $\mathcal{J}_{\pm I}$, Eqs. (\ref{Eq:JIpm}), (\ref{Eq:JpmI}):\\
 Using Eq. (\ref{Eq:Jcc}), 
\begin{eqnarray}
 \mathcal{J}_{I\pm} &=& \sum_{kk'}^{\rm even} 
                        \mathcal{J}_{k0k'\pm 2m_2} 
                        \frac{C_{J_1m_1k0}^{J_1m_1} C_{J_2-m_2k'2m_2}^{J_2m_2}}
                        {C_{J_1J_1 k0}^{J_1J_1} C_{J_2J_2 k'0}^{J_2J_2}}
\nonumber\\
                    &=& (-1)^{\pm 2m_2} 
                        \sum_{kk'}^{\rm even} 
                        \left(\mathcal{J}_{k0k'\mp 2m_2}\right)^*
                        \frac{C_{J_1m_1k0}^{J_1m_1} C_{J_2-m_2k'2m_2}^{J_2m_2}}
                        {C_{J_1J_1 k0}^{J_1J_1} C_{J_2J_2 k'0}^{J_2J_2}}
\nonumber\\
                    &=& (-1)^{2m_2} \mathcal{J}_{I\mp}^*.
\end{eqnarray}
 Since $m_2$ is integer or half-integer, $(-1)^{2m_2}=(-1)^{-2m_2}$.
 On the other hand, applying Eq. (\ref{Eq:CG_2}) to site 2, 
\begin{eqnarray}
 \mathcal{J}_{I\pm} &=& \sum_{kk'}^{\rm even} (-1)^{k'+2m_2}
                        \mathcal{J}_{k0k'\pm 2m_2} 
                        \frac{C_{J_1m_1k0}^{J_1m_1} C_{J_2-m_2k'2m_2}^{J_2m_2}}
                        {C_{J_1J_1 k0}^{J_1J_1} C_{J_2J_2 k'0}^{J_2J_2}}
\nonumber\\
                    &=& (-1)^{2m_2} \mathcal{J}_{I\pm}.
\end{eqnarray}
 Since $k'$ is even, $(-1)^{k'}=1$. Therefore, 
\begin{eqnarray}
 \mathcal{J}_{I\pm} = (-1)^{2m_2} \mathcal{J}_{I\pm} = \mathcal{J}_{I\mp}^*,
\end{eqnarray}
 and we obtain Eq. (\ref{Eq:JIpm}).
 Similarly, we obtain Eq. (\ref{Eq:JpmI}).
 \item $\mathcal{J}_{\pm \pm}$, Eq. (\ref{Eq:Jpmpm}):\\
 Using Eq. (\ref{Eq:Jcc}), 
\begin{eqnarray}
 \mathcal{J}_{\pm \pm} &=& \sum_{kk'}^{\rm even} \mathcal{J}_{k\pm 2m_1 k'\pm 2m_2} 
                           \frac{C_{J_1-m_1 k 2m_1}^{J_1m_1} C_{J_2-m_2 k'2m_2}^{J_2m_2}}
                           {C_{J_1J_1 k0}^{J_1J_1} C_{J_2J_2 k'0}^{J_2J_2}}
\nonumber\\ 
                       &=& (-1)^{2(m_1+m_2)} 
                           \sum_{kk'}^{\rm even} \left(\mathcal{J}_{k\mp 2m_1 k'\mp 2m_2}\right)^*
                           \frac{C_{J_1-m_1 k 2m_1}^{J_1m_1} C_{J_2-m_2 k'2m_2}^{J_2m_2}}
                           {C_{J_1J_1 k0}^{J_1J_1} C_{J_2J_2 k'0}^{J_2J_2}}
\nonumber\\ 
                       &=& (-1)^{2(m_1+m_2)} \mathcal{J}_{\mp \mp}^*.
\end{eqnarray}
 Using Eq. (\ref{Eq:CG_2}) for both sites, 
\begin{eqnarray}
 \mathcal{J}_{\pm \pm} &=& 
                           \sum_{kk'}^{\rm even} 
                           (-1)^{k+2m_1+k'+2m_2} 
                           \mathcal{J}_{k\pm 2m_1 k'\pm 2m_2} 
                           \frac{C_{J_1-m_1 k 2m_1}^{J_1m_1} C_{J_2-m_2 k'2m_2}^{J_2m_2}}
                           {C_{J_1J_1 k0}^{J_1J_1} C_{J_2J_2 k'0}^{J_2J_2}}
\nonumber\\ 
                       &=& (-1)^{2(m_1+m_2)} \mathcal{J}_{\pm \pm}.
\end{eqnarray}
% In the last equation, we use the fact that both $k$ and $k'$ are even.
 From these equations, 
\begin{eqnarray}
 \mathcal{J}_{\pm \pm} = (-1)^{2(m_1+m_2)} \mathcal{J}_{\pm \pm}
                       = (-1)^{2(m_1+m_2)} \mathcal{J}_{\mp \mp}^*,
\end{eqnarray}
 and we obtain relation (\ref{Eq:Jpmpm}).
 \item $\mathcal{J}_{\pm \mp}$, Eq. (\ref{Eq:Jpmmp}):\\
 Using Eq. (\ref{Eq:Jcc}), 
\begin{eqnarray}
 \mathcal{J}_{\pm \mp} &=& \sum_{kk'}^{\rm even} \mathcal{J}_{k\pm 2m_1 k'\mp 2m_2} 
                           \frac{C_{J_1-m_1 k 2m_1}^{J_1m_1} C_{J_2-m_2 k'2m_2}^{J_2m_2}}
                           {C_{J_1J_1 k0}^{J_1J_1} C_{J_2J_2 k'0}^{J_2J_2}}
\nonumber\\
                       &=& (-1)^{2(m_1 + m_2)}
                           \sum_{kk'}^{\rm even} 
                           \left(\mathcal{J}_{k\mp 2m_1 k'\pm 2m_2}\right)^*
                           \frac{C_{J_1-m_1 k 2m_1}^{J_1m_1} C_{J_2-m_2 k'2m_2}^{J_2m_2}}
                           {C_{J_1J_1 k0}^{J_1J_1} C_{J_2J_2 k'0}^{J_2J_2}}
\nonumber\\
                       &=& (-1)^{2m_1 + 2m_2} \mathcal{J}_{\mp \pm}^*.
\end{eqnarray}
 Using Eq. (\ref{Eq:CG_2}) for both sites, 
\begin{eqnarray}
 \mathcal{J}_{\pm \mp} &=& \sum_{kk'}^{\rm even} (-1)^{k+2m_1+k'+2m_2} 
                           \mathcal{J}_{k\pm 2m_1 k'\mp 2m_2} 
                           \frac{C_{J_1-m_1 k 2m_1}^{J_1m_1} C_{J_2-m_2 k'2m_2}^{J_2m_2}}
                           {C_{J_1J_1 k0}^{J_1J_1} C_{J_2J_2 k'0}^{J_2J_2}}
\nonumber\\
                       &=& (-1)^{2(m_1+m_2)} \mathcal{J}_{\pm \mp}.
\end{eqnarray}
 Combinig these relations, 
\begin{eqnarray}
 \mathcal{J}_{\pm \mp} = (-1)^{2(m_1+m_2)} \mathcal{J}_{\pm \mp}
                       = (-1)^{2(m_1 + m_2)} \mathcal{J}_{\mp \pm}^*,
\end{eqnarray}
 and we obtain Eq. (\ref{Eq:Jpmmp}).
 \item $\mathcal{K}_{z\pm}$, $\mathcal{K}_{\pm z}$, Eqs. (\ref{Eq:Kzpm}), (\ref{Eq:Kpmz}):\\
 Using Eq. (\ref{Eq:Jcc}), 
\begin{eqnarray}
 \mathcal{K}_{z\pm} &=& \pm \sum_{kk'}^{\rm odd} \mathcal{J}_{k0k'\pm 2m_2} 
                        \frac{C_{J_1m_1k0}^{J_1m_1} C_{J_2-m_2 k'2m_2}^{J_2m_2}}
                        {C_{J_1J_1k0}^{J_1J_1} C_{J_2J_2 k'0}^{J_2J_2}}
\nonumber\\
                    &=& \pm (-1)^{2m_2} \sum_{kk'}^{\rm odd} \left(\mathcal{J}_{k0k'\mp 2m_2}\right)^*
                        \frac{C_{J_1m_1k0}^{J_1m_1} C_{J_2-m_2 k'2m_2}^{J_2m_2}}
                        {C_{J_1J_1k0}^{J_1J_1} C_{J_2J_2 k'0}^{J_2J_2}}
\nonumber\\
                    &=& -(-1)^{2m_2}\mathcal{K}_{z\mp}^*.
\end{eqnarray}
 Using Eq. (\ref{Eq:CG_2}) for site 2, 
\begin{eqnarray}
 \mathcal{K}_{z\pm} &=& \pm \sum_{kk'}^{\rm odd} 
                        (-1)^{k'+2m_2} \mathcal{J}_{k0k'\pm 2m_2} 
                        \frac{C_{J_1m_1k0}^{J_1m_1} C_{J_2-m_2 k'2m_2}^{J_2m_2}}
                        {C_{J_1J_1k0}^{J_1J_1} C_{J_2J_2 k'0}^{J_2J_2}}
\nonumber\\
                    &=& -(-1)^{2m_2} \mathcal{K}_{z\pm}.
\end{eqnarray}
 Here, we used that $k'$ is odd. 
 Therefore, we obtain 
\begin{eqnarray}
 \mathcal{K}_{z\pm} = -(-1)^{2m_2} \mathcal{K}_{z\pm}
                    = -(-1)^{2m_2} \mathcal{K}_{z\mp}^*,
\end{eqnarray}
 and Eq. (\ref{Eq:Kzpm}).
 Similarly, we obtain Eq, (\ref{Eq:Kpmz}).
 \item $\mathcal{K}_{\pm\pm}$, Eq. (\ref{Eq:Kpmpm}):\\
 Using Eq. (\ref{Eq:Jcc}), 
\begin{eqnarray}
 \mathcal{K}_{\pm \pm} &=& \sum_{kk'}^{\rm odd} \mathcal{J}_{k\pm 2m_1 k'\pm 2m_2}
                           \frac{C_{J_1-m_1k2m_1}^{J_1m_1} C_{J_2-m_2k'2m_2}^{J_2m_2}}
                           {C_{J_1J_1k0}^{J_1J_1} C_{J_2J_2k'0}^{J_2J_2}}
\nonumber\\
                       &=& (-1)^{2(m_1+m_2)} 
                           \sum_{kk'}^{\rm odd} \left(\mathcal{J}_{k\mp 2m_1 k'\mp 2m_2}\right)^*
                           \frac{C_{J_1-m_1k2m_1}^{J_1m_1} C_{J_2-m_2k'2m_2}^{J_2m_2}}
                           {C_{J_1J_1k0}^{J_1J_1} C_{J_2J_2k'0}^{J_2J_2}}
\nonumber\\
                       &=& (-1)^{2(m_1+m_2)} \mathcal{K}_{\mp \mp}^*.
\end{eqnarray}
 Using Eq. (\ref{Eq:CG_2}) for both sites, 
\begin{eqnarray}
 \mathcal{K}_{\pm \pm} &=& \sum_{kk'}^{\rm odd} (-1)^{k+2m_1 + k'+2m_2}
                           \mathcal{J}_{k\pm 2m_1 k'\pm 2m_2}
                           \frac{C_{J_1-m_1k2m_1}^{J_1m_1} C_{J_2-m_2k'2m_2}^{J_2m_2}}
                           {C_{J_1J_1k0}^{J_1J_1} C_{J_2J_2k'0}^{J_2J_2}}
\nonumber\\
%                       &=& (-1)^{2m_1 +2m_2}\sum_{kk'}^{\rm odd} 
%                           \mathcal{J}_{k\pm 2m_1 k'\pm 2m_2}
%                           \frac{C_{J_1-m_1k2m_1}^{J_1m_1} C_{J_2-m_2k'2m_2}^{J_2m_2}}
%                           {C_{J_1J_1k0}^{J_1J_1} C_{J_2J_2k'0}^{J_2J_2}}
                       &=& (-1)^{2(m_1 + m_2)} \mathcal{K}_{\pm \pm}.
\end{eqnarray}
 Therefore, 
\begin{eqnarray}
 \mathcal{K}_{\pm \pm} = (-1)^{2(m_1+m_2)} \mathcal{K}_{\pm \pm}
                       = (-1)^{2(m_1+m_2)} \mathcal{K}_{\mp \mp}^*.
\end{eqnarray}
 \item $\mathcal{K}_{\pm \mp}$, Eq. (\ref{Eq:Kpmmp}):\\
 Using Eq. (\ref{Eq:Jcc}), 
\begin{eqnarray}
 \mathcal{K}_{\pm \mp} &=& -\sum_{kk'}^{\rm odd} \mathcal{J}_{k\pm 2m_1 k'\mp 2m_2}
                           \frac{C_{J_1-m_1k2m_1}^{J_1m_1} C_{J_2-m_2k'2m_2}^{J_2m_2}}
                           {C_{J_1J_1k0}^{J_1J_1} C_{J_2J_2k'0}^{J_2J_2}}
\nonumber\\
                       &=& -(-1)^{2(m_1+m_2)}\sum_{kk'}^{\rm odd} \left(\mathcal{J}_{k\mp 2m_1 k'\pm 2m_2}\right)^*
                           \frac{C_{J_1-m_1k2m_1}^{J_1m_1} C_{J_2-m_2k'2m_2}^{J_2m_2}}
                           {C_{J_1J_1k0}^{J_1J_1} C_{J_2J_2k'0}^{J_2J_2}}
\nonumber\\
                       &=& (-1)^{2m(m_1+m_2)} \mathcal{K}_{\mp \pm}^*.
\end{eqnarray}
 Using Eq. (\ref{Eq:CG_2}) for both sites,  
\begin{eqnarray}
 \mathcal{K}_{\pm \mp} &=& -\sum_{kk'}^{\rm odd} (-1)^{k+2m_1+k'+2m_2}\mathcal{J}_{k\pm 2m_1 k'\mp 2m_2}
                           \frac{C_{J_1-m_1k2m_1}^{J_1m_1} C_{J_2-m_2k'2m_2}^{J_2m_2}}
                           {C_{J_1J_1k0}^{J_1J_1} C_{J_2J_2k'0}^{J_2J_2}}
\nonumber\\
                       &=& (-1)^{2m_1+2m_2} \mathcal{K}_{\pm \mp}.
\end{eqnarray}
 Therefore, 
\begin{eqnarray}
 \mathcal{K}_{\pm \mp} = (-1)^{2(m_1+m_2)} \mathcal{K}_{\pm \mp} = (-1)^{2(m_1+m_2)} \mathcal{K}_{\mp \pm}^*,
\end{eqnarray}
and we obtain Eq. (\ref{Eq:Kpmmp}).
\end{itemize}

\subsection{Symmetry properties}
\label{A:JK}
Some of $\mathcal{J}$ and $\mathcal{K}$ become zero due to (i) time reversal symmetry 
and (ii) the structure of the exchange interaction.
\begin{enumerate}
 \item Time reversal symmetry:
\\
When magnetic site has odd (even) number of electrons, 
the total angular momentum and its projection $m_i$ are half-integer (integer),
and thus, $(-1)^{2m_i} = -1$ $(=1)$.
Due to the change in sign, some of $\mathcal{J}$ and $\mathcal{K}$ (\ref{Eq:JII})-(\ref{Eq:Kpmmp}) become zero.
The results are summarized as follows:
%\sout{
% Due to the time reversal symmetry of the magnetic sites, some of $\mathcal{J}$ and $\mathcal{K}$ become zero.
%}
\begin{itemize}
 \item $\mathcal{J}_{I\pm} = 0$ when $N_2$ is odd,
 \item $\mathcal{J}_{\pm I} = 0$ when $N_1$ is odd, 
 \item $\mathcal{K}_{z\pm} = 0$ when $N_2$ is even,
 \item $\mathcal{K}_{\pm z} = 0$ when $N_1$ is even, 
% \item $\mathcal{J}_{\pm\pm} = \mathcal{J}_{\pm \mp} = \mathcal{K}_{\pm \pm} = \mathcal{K}_{\mp \mp} = 0$ when $N_1 + N_2$ is odd,
 \item $\mathcal{J}_{\pm \pm} = \mathcal{J}_{\pm \mp}=0$ when 
%\sout{at least} 
one of $N_1$ and $N_2$ is odd,
 \item $\mathcal{K}_{\pm \pm} = \mathcal{K}_{\pm \mp}=0$ when 
%\sout{at least} 
one of $N_1$ and $N_2$ is even,
\end{itemize}
where $N_1$ and $N_2$ are the numbers of electrons on sites $1$ and $2$, respectively.
 \item %\sout{Relation between $m_i$ and $q_{\rm max}$.}
 Structure of the exchange Hamiltonian:
\\
% The structure of the effective Hamiltonian is determined by the relation between $m_i$ and $q_{\rm max}$.
 The values of the exchange parameters $\mathcal{J}$ and $\mathcal{K}$ depend on $q_{\rm max}$ (\ref{Eq:qmax1}).
 For large $m_i > q_{\rm max}/2$, some of $\mathcal{J}$ and $\mathcal{K}$ are zero because
 either Clebsch-Gordan coefficient $C_{J-mk2m}^{Jm}$ or $\mathcal{J}_{kqk'q'}$ becomes zero.
\\ 

 First, we consider non-symmetric system where $q_{\rm max}$ corresponds to $k_{\rm max}$.
 In this case, 
%\sout{the condition that} 
 $\mathcal{J}$ and $\mathcal{K}$ become zero 
%\sout{owing to the value of the Clebsch-Gordan coefficients} 
only when $C_{J-mk2m}^{Jm}=0$, which holds when $k>2m$.
Since maximum of $k$ is fixed, we have to further consider two cases regarding $J$:
 \begin{enumerate}
% \item $k_{\rm max}, q_{\rm max} = 2J_i < 2(l_i + 1/2)$ ($N=1,5$)
%    or $k_{\rm max}, q_{\rm max} = 2J_i = 2(l_i + 1/2)$ ($N=7,13$):
  \item $J_i \le l_i+1/2$ and $k_{\rm max}=q_{\rm max}=2J_i$:
\\
  Since $2m \le k_{\rm max}$ is always satisfied, 
  Clebsch-Gordan coefficients are in general nonzero,
  $C_{J-mk_{\rm max} 2m}^{Jm} \ne 0$.
  Therefore, all $\mathcal{J}$ and $\mathcal{K}$ can be nonzero
%\sout{
%  (Here, time reversal symmetry is not taken into account).
%}
  due to the reason different from time reversal symmetry.
 \item $J_i > l_i+1/2$ and $k_{\rm max} = q_{\rm max} = 2(l_i + 1/2)$:
\\
%\sout{
%   Choosing $2m > k_{\rm max}$, 
%}
    It is possible to have $m$ which is larger than $k_{\rm max}/2$. 
    In this case, we obtain
    $C_{J-mk_{\rm max}2m}^{Jm} = 0$, and $\mathcal{J}$ and $\mathcal{K}$ which include
    the Clebsch-Gordan coefficient become zero.
 \end{enumerate}
% Here, $N$ in the parenthesis is the number of $f$ electrons.  
 The case of $N=6$ ($J=0$) is not considered since the ion is nonmagnetic.
%\sout{
% Cases (a), (b) are summarized as 
%}
 The above discussion is summarized as follows:
 \begin{itemize}
  \item $\mathcal{J}_{\pm I} = \mathcal{J}_{\pm \pm} = \mathcal{J}_{\pm \mp} = 
         \mathcal{K}_{\pm z} = \mathcal{K}_{\pm \pm} = \mathcal{K}_{\pm \mp} =0$ when $m_1 > l_1+1/2$,
  \item $\mathcal{J}_{I \pm} = \mathcal{J}_{\pm \pm} = \mathcal{J}_{\pm \mp} =
         \mathcal{K}_{z \pm} = \mathcal{K}_{\pm \pm} = \mathcal{K}_{\pm \mp} =0$ when $m_2 > l_2+1/2$.
 \end{itemize}

 In symmetric systems, 
 $q_{\rm max} \le k_{\rm max}$, and the condition changes.
%\sout{
% the values of $\mathcal{J}$ and $\mathcal{K}$ 
% depends on both $\mathcal{J}_{kqk'q'}$ and the Clebsch-Gordan coefficients. 
% For example, if there is }
 Contrary to the non-symmetric case, $\mathcal{J}$ and $\mathcal{K}$ become zero for 
 $m$ such that $k_{\rm max} > 2m > q_{\rm max} \ge 0$,
 because $C_{J-mk2m}^{Jm}$ is not zero for $k \ge 2m$, while $\mathcal{J}_{k2mk'q'} = 0$. 
%\sout{
%, and hence their product is also zero.
%}
 The selection rule becomes
 \begin{itemize}
  \item $\mathcal{J}_{\pm I} = \mathcal{J}_{\pm \pm} = \mathcal{J}_{\pm \mp} = 
         \mathcal{K}_{\pm z} = \mathcal{K}_{\pm \pm} = \mathcal{K}_{\pm \mp} =0$ when $m_1 > q_{\rm max}/2$,
  \item $\mathcal{J}_{I \pm} = \mathcal{J}_{\pm \pm} = \mathcal{J}_{\pm \mp} = 
         \mathcal{K}_{z \pm} = \mathcal{K}_{\pm \pm} = \mathcal{K}_{\pm \mp} =0$ when $m_2 > q'_{\rm max}/2$.
 \end{itemize}
% Note that in general $q_{\rm max} \le k_{\rm max}$ ($q'_{\rm max} \le k'_{\rm max}$),
% if there are $m_1$ ($m_2$) such that $k_{\rm max} \ge 2m_1 > q_{\rm max}$ 
% ($k'_{\rm max} \ge 2m_2 > q'_{\rm max}$), 
% some $\mathcal{J}$ and $\mathcal{K}$ are zero even in the case of $2l_1 + 1 \ge k_{\rm max}$
% ($2l_2 + 1 \ge k'_{\rm max}$).
 %For example, consider symmetric Yb dimer complex with $q_{\rm max} = 5$ and $m = 7/2$,
 %all $\mathcal{K}$ mentioned here are zero ($\mathcal{J}$ are zero because of time-reversal symmetry).
\end{enumerate}

%\subsection{Exchange Hamiltonian fo $J$-multiplet and isotropic spin}
%In 

\section{Classification of the effective Hamiltonians}
\label{A:type_Ising}
With the use of the conditions given in Sec. \ref{A:JK}, 
we study the structure of the effective Hamiltonian, (\ref{Eq:H_Lines_2}) and (\ref{Eq:H_doub_iso}), for all cases.
The effective Hamiltonian is classified into three types, Ising I, Ising II, and non-Ising.
Ising I does not change the magnetic axes due to the exchange interaction, while Ising II does. 
%The result is summarized in Table \ref{Table:Heff}.

%\begin{table}[tbh]
%\caption{Classification of the effective Hamiltonian. 
%``odd'' and ``even'' indicate the number of the electrons, and
%``I'', ``II'' the type of the Ising model and ``N'' the non-Ising.
%In the case of type I, the direction of the magnetization axis 
%does not vary due to the exchange interaction.
%In general, $q_{\rm max}$ (\ref{Eq:qmax1}) and $q_{\rm max}'$ (\ref{Eq:qmax2}) 
%depend on the geometry of the bridging ligand.
%When the structure of the bridging ligand does not have high symmetry, 
%$q_{\rm max}$ ($q_{\rm max}'$) in this table is replaced by $l_1+1/2$ ($l_2+1/2$).
%}
%\label{Table:Heff}
%\begin{ruledtabular}
%\begin{tabular}{cccccc}
%\multirow{2}{*}{\backslashbox{$N_2$}{$N_1$}} 
%&& \multicolumn{2}{c}{odd} & \multicolumn{2}{c}{even} \\
%&& $m_1 > q_{\rm max}/2$ & $m_1 \le q_{\rm max}/2$ & $m_1> q_{\rm max}$ & $m_1 \le q_{\rm max}$ \\
%&& $m_1 > l_1+\frac{1}{2}$ & $m_1 \le l_1+\frac{1}{2}$ & $m_1> l_1+\frac{1}{2}$ & $m_1< l_1+\frac{1}{2}$ \\
%\hline
%odd  & $m_2 >   q'_{\rm max}/2$   & I  & II & I  & N \\
%odd  & $m_2 \le q'_{\rm max}/2$   & II & N  & II & N \\
%even & $m_2 >   q'_{\rm max}/2$   & I  & II & I  & N \\
%even & $m_2 \le q'_{\rm max}/2$   & N  & N  & N  & N \\
%odd  & $m_2 >   l_2+\frac{1}{2}$   & I  & II & I  & N \\
%odd  & $m_2 \le l_2+\frac{1}{2}$   & II & N  & II & N \\
%even & $m_2 >   l_2+\frac{1}{2}$   & I  & II & I  & N \\
%even & $m_2 <   l_2+\frac{1}{2}$   & N  & N  & N  & N \\
%\multicolumn{2}{c}{isotropic spin} & II & N  & II & N \\
%\end{tabular}
%\end{ruledtabular}
%\end{table}

\subsection{Doublets with unquenched orbital momentum}
The form of the effective Hamiltonian 
between anisotropic doublets (\ref{Eq:H_Lines_2}) becomes as follows
(the condition for non-symmetric case is obtained by replacing $q_{\rm max}$ by $l+1/2$):
\begin{enumerate}
 \item Both of $N_1$ and $N_2$ are odd ($m_1$ and $m_2$ are half-integer).
 \begin{enumerate}
 \item $m_1 > q_{\rm max}/2$ and $m_2 > q'_{\rm max}/2$ (Ising, I)
\begin{eqnarray}
 \hat{\rm H}_{\rm eff} &=&
   \mathcal{J}_{II} \tilde{I}_1 \tilde{I}_2 
 + \mathcal{K}_{zz} \frac{\tilde{s}_{1z_1}}{\tilde{s}_1} \frac{\tilde{s}_{2z_2}}{\tilde{s}_2}.
\end{eqnarray}
 \item $m_1 > q_{\rm max}/2$ and $m_2 \le q'_{\rm max}/2$ (Ising, II)
% \item $m_1 > l_1 + \frac{1}{2}$ and $m_2 \le l_2 + \frac{1}{2}$ (Ising, II)
\begin{eqnarray}
 \hat{\rm H}_{\rm eff} &=&
   \mathcal{J}_{II} \tilde{I}_1 \tilde{I}_2 
 + \mathcal{K}_{zz} \frac{\tilde{s}_{1z_1}}{\tilde{s}_1} \frac{\tilde{s}_{2z_2}}{\tilde{s}_2}
 + \mathcal{K}_{z+} \frac{\tilde{s}_{1z_1}}{\tilde{s}_1} \tilde{s}_{2+} 
 + \mathcal{K}_{z-} \frac{\tilde{s}_{1z_1}}{\tilde{s}_1} \tilde{s}_{2-}
\\
 &=&
   \mathcal{J}_{II} \tilde{I}_1 \tilde{I}_2 
 + \mathcal{K}'_{zz'} \frac{\tilde{s}_{1z_1}}{\tilde{s}_1} \frac{\tilde{s}_{2z'}}{\tilde{s}_2},
\end{eqnarray}
where
\begin{eqnarray}
 \mathcal{K}'_{zz'} &=& \sqrt{\mathcal{K}_{zz}^2+\left|\mathcal{K}_{z+}\right|^2},
\\
 \tilde{s}_{2z'} &=& 
   \frac{\Re(\mathcal{K}_{z+})}{\mathcal{K}'_{zz'}}\tilde{s}_{2x_2} 
 - \frac{\Im(\mathcal{K}_{z+})}{\mathcal{K}'_{zz'}}\tilde{s}_{2y_2}
 + \frac{\mathcal{K}_{zz}}{\mathcal{K}'_{zz'}}\tilde{s}_{2z_2}.
\end{eqnarray}
 \item $m_1 \le q_{\rm max}/2$ and $m_2 > q'_{\rm max}/2$ (Ising, II)
% \item $m_1 \le l_1+\frac{1}{2}$ and $m_2 > l_2+\frac{1}{2}$ (Ising, II)
\begin{eqnarray}
 \hat{\rm H}_{\rm eff} &=&
   \mathcal{J}_{II} \tilde{I}_1 \tilde{I}_2 
 + \mathcal{K}_{zz} \frac{\tilde{s}_{1z_1}}{\tilde{s}_1} \frac{\tilde{s}_{2z_2}}{\tilde{s}_2}
 + \mathcal{K}_{+z} \tilde{s}_{1+} \frac{\tilde{s}_{2z_2}}{\tilde{s}_2}
 + \mathcal{K}_{-z} \tilde{s}_{1-} \frac{\tilde{s}_{2z_2}}{\tilde{s}_2}
\\
 &=&
   \mathcal{J}_{II} \tilde{I}_1 \tilde{I}_2 
 + \mathcal{K}'_{z'z} \frac{\tilde{s}_{1z'}}{\tilde{s}_1} \frac{\tilde{s}_{2z_2}}{\tilde{s}_2},
% + \mathcal{K}_{+z} \tilde{s}_{1+} \frac{\tilde{s}_{2z}}{\tilde{s}_2}
% + \mathcal{K}_{-z} \tilde{s}_{1-} \frac{\tilde{s}_{2z}}{\tilde{s}_2}
\label{Eq:H_Lines_23}
\end{eqnarray}
where
\begin{eqnarray}
 \mathcal{K}'_{zz} &=& \sqrt{\mathcal{K}_{zz}^2+\left|\mathcal{K}_{+z}\right|^2},
\\
 \tilde{s}_{1z'} &=& 
   \frac{\Re(\mathcal{K}_{+z})}{\mathcal{K}'_{z'z}}\tilde{s}_{1x_1} 
 - \frac{\Im(\mathcal{K}_{+z})}{\mathcal{K}'_{z'z}}\tilde{s}_{1y_1}
 + \frac{\mathcal{K}_{zz}}{\mathcal{K}'_{z'z}}\tilde{s}_{1z_1}.
\end{eqnarray}
 \item $m_1 \le q_{\rm max}/2$ and $m_2 \le q'_{\rm max}/2$ (non-Ising)
% \item $m_1 \le l_1+\frac{1}{2}$ and $m_2 \le l_2+\frac{1}{2}$ (non-Ising)
\begin{eqnarray}
 \hat{\rm H}_{\rm eff} &=&
   \mathcal{J}_{II} \tilde{I}_1 \tilde{I}_2 
 + \mathcal{K}_{zz} \frac{\tilde{s}_{1z_1}}{\tilde{s}_1} \frac{\tilde{s}_{2z_2}}{\tilde{s}_2}
\nonumber\\
 &+&
   \mathcal{K}_{z+} \frac{\tilde{s}_{1z_1}}{\tilde{s}_1} \tilde{s}_{2+} 
 + \mathcal{K}_{z-} \frac{\tilde{s}_{1z_1}}{\tilde{s}_1} \tilde{s}_{2-} 
\nonumber\\
 &+&
   \mathcal{K}_{+z} \tilde{s}_{1+} \frac{\tilde{s}_{2z_2}}{\tilde{s}_2}
 + \mathcal{K}_{-z} \tilde{s}_{1-} \frac{\tilde{s}_{2z_2}}{\tilde{s}_2}
\nonumber\\
 &+&
   \mathcal{K}_{++} \tilde{s}_{1+} \tilde{s}_{2+} 
 + \mathcal{K}_{+-} \tilde{s}_{1+} \tilde{s}_{2-} 
\nonumber\\
 &+&
   \mathcal{K}_{-+} \tilde{s}_{1-} \tilde{s}_{2+} 
 + \mathcal{K}_{--} \tilde{s}_{1-} \tilde{s}_{2-}
\\
 &=&
   \mathcal{J}_{II} \tilde{I}_1 \tilde{I}_2 
 + \mathcal{K}_{zz} \frac{\tilde{s}_{1z_1}}{\tilde{s}_1} \frac{\tilde{s}_{2z_2}}{\tilde{s}_2}
\nonumber\\ 
 &+&
   \Re(\mathcal{K}_{z+}) \frac{\tilde{s}_{1z_1}\tilde{s}_{2x_2}}{\tilde{s}_1\tilde{s}_2} 
 - \Im(\mathcal{K}_{z+}) \frac{\tilde{s}_{1z_1}\tilde{s}_{2y_2}}{\tilde{s}_1\tilde{s}_2} 
\nonumber\\
 &+&
   \Re(\mathcal{K}_{+z}) \frac{\tilde{s}_{1x_1}\tilde{s}_{2z_2}}{\tilde{s}_1\tilde{s}_2}
 - \Im(\mathcal{K}_{+z}) \frac{\tilde{s}_{1y_1}\tilde{s}_{2z_2}}{\tilde{s}_1\tilde{s}_2}
\nonumber\\
 &+&
   \frac{\Re(\mathcal{K}_{++})}{2}
   \frac{\tilde{s}_{1x_1} \tilde{s}_{2x_2} - \tilde{s}_{1y_1} \tilde{s}_{2y_2}}{\tilde{s}_1\tilde{s}_2}
\nonumber\\
 &-&
   \frac{\Im(\mathcal{K}_{++})}{2} 
   \frac{\tilde{s}_{1x_1} \tilde{s}_{2y_2} + \tilde{s}_{1y_1} \tilde{s}_{2x_2}}{\tilde{s}_1\tilde{s}_2}
\nonumber\\
 &+&
   \frac{\Re(\mathcal{K}_{+-})}{2} 
   \frac{\tilde{s}_{1x_1} \tilde{s}_{2x_2} + \tilde{s}_{1y_1} \tilde{s}_{2y_2}}{\tilde{s}_1\tilde{s}_2}
\nonumber\\
 &-&
   \frac{\Im(\mathcal{K}_{+-})}{2} 
   \frac{\tilde{s}_{1x_1} \tilde{s}_{2y_2} - \tilde{s}_{1y_1} \tilde{s}_{2x_2}}{\tilde{s}_1\tilde{s}_2}.
\label{Eq:H_Lines_24}
\end{eqnarray}
 \end{enumerate}
 \item $N_1$ is even and $N_2$ is odd ($m_1$ is integer, $m_2$ is half-integer).
 \begin{enumerate}
  \item $m_1 > q_{\rm max}/2$ and $m_2 > q'_{\rm max}/2$ (Ising, I)
% \item $m_1 > l_1+\frac{1}{2}$ and $m_2 > l_2+\frac{1}{2}$ (Ising, I)
\begin{eqnarray}
 \hat{\rm H}_{\rm eff} &=&
   \mathcal{J}_{II} \tilde{I}_1 \tilde{I}_2 
 + \mathcal{K}_{zz} \frac{\tilde{s}_{1z_1}}{\tilde{s}_1} \frac{\tilde{s}_{2z_2}}{\tilde{s}_2}.
\label{Eq:H_Lines_11}
\end{eqnarray}
 \item $m_1 > q_{\rm max}/2$ and $m_2 \le q'_{\rm max}/2$ (Ising, II)
% \item $m_1 > l_1+\frac{1}{2}$ and $m_2 \le l_2+\frac{1}{2}$ (Ising, II)
\begin{eqnarray}
 \hat{\rm H}_{\rm eff} &=&
   \mathcal{J}_{II} \tilde{I}_1 \tilde{I}_2 
 + \mathcal{K}_{zz} \frac{\tilde{s}_{1z_1}}{\tilde{s}_1} \frac{\tilde{s}_{2z_2}}{\tilde{s}_2}
 + \mathcal{K}_{z+} \frac{\tilde{s}_{1z_1}}{\tilde{s}_1} \tilde{s}_{2+} 
 + \mathcal{K}_{z-} \frac{\tilde{s}_{1z_1}}{\tilde{s}_1} \tilde{s}_{2-}
\\
 &=&
   \mathcal{J}_{II} \tilde{I}_1 \tilde{I}_2 
 + \mathcal{K}'_{zz'} \frac{\tilde{s}_{1z_1}}{\tilde{s}_1} \frac{\tilde{s}_{2z'}}{\tilde{s}_2},
\label{Eq:H_Lines_12}
\end{eqnarray}
where
\begin{eqnarray}
 \mathcal{K}'_{zz'} &=& \sqrt{\mathcal{K}_{zz}^2 + \left|\mathcal{K}_{z+}\right|^2},
\\
 \tilde{s}_{2z'}
 &=&
   \frac{\Re(\mathcal{K}_{z+})}{\mathcal{K}'_{zz'}} \tilde{s}_{2x_2}
 - \frac{\Im(\mathcal{K}_{z+})}{\mathcal{K}'_{zz'}} \tilde{s}_{2y_2}
 + \frac{\mathcal{K}_{zz}}{\mathcal{K}'_{zz'}} \tilde{s}_{2z_2}.
\end{eqnarray}
 \item $m_1 \le q_{\rm max}/2$ and $m_2 > q'_{\rm max}/2$ (non-Ising)
% \item $m_1 < l_1+\frac{1}{2}$ and $m_2 > l_2+\frac{1}{2}$ (non-Ising)
\begin{eqnarray}
 \hat{\rm H}_{\rm eff} &=&
   \mathcal{J}_{II} \tilde{I}_1 \tilde{I}_2 
 + \mathcal{K}_{zz} \frac{\tilde{s}_{1z_1}}{\tilde{s}_1} \frac{\tilde{s}_{2z_2}}{\tilde{s}_2}
 + \mathcal{J}_{+I} \tilde{s}_{1+} \tilde{I}_{2} 
 + \mathcal{J}_{-I} \tilde{s}_{1-} \tilde{I}_{2}.
\label{Eq:H_Lines_13}
\end{eqnarray}
 \item $m_1 \le q_{\rm max}/2$ and $m_2 \le q'_{\rm max}/2$ (non-Ising)
% \item $m_1 < l_1+\frac{1}{2}$ and $m_2 \le l_2+\frac{1}{2}$ (non-Ising)
\begin{eqnarray}
 \hat{\rm H}_{\rm eff} &=&
   \mathcal{J}_{II} \tilde{I}_1 \tilde{I}_2 
 + \mathcal{K}_{zz} \frac{\tilde{s}_{1z_1}}{\tilde{s}_1} \frac{\tilde{s}_{2z_2}}{\tilde{s}_2}
\nonumber\\
 &+&
   \mathcal{J}_{+I} \tilde{s}_{1+} \tilde{I}_{2} 
 + \mathcal{J}_{-I} \tilde{s}_{1-} \tilde{I}_{2} 
   \mathcal{K}_{z+} \frac{\tilde{s}_{1z_1}}{\tilde{s}_1} \tilde{s}_{2+} 
 + \mathcal{K}_{z-} \frac{\tilde{s}_{1z_1}}{\tilde{s}_1} \tilde{s}_{2-}.
\label{Eq:H_Lines_14}
\end{eqnarray}
\end{enumerate}
 \item Both of $N_1$ and $N_2$ are even ($m_1$ and $m_2$ are integer).
 \begin{enumerate}
 \item $m_1 > q_{\rm max}/2$ and $m_2 > q'_{\rm max}/2$ (Ising, I)
% \item $m_1 > l_1+\frac{1}{2}$ and $m_2 > l_2+\frac{1}{2}$ (Ising, I)
\begin{eqnarray}
 \hat{\rm H}_{\rm eff} &=&
   \mathcal{J}_{II} \tilde{I}_1 \tilde{I}_2 
 + \mathcal{K}_{zz} \frac{\tilde{s}_{1z_1}}{\tilde{s}_1} \frac{\tilde{s}_{2z_2}}{\tilde{s}_2}.
\label{Eq:H_Lines_C1}
\end{eqnarray}
 \item $m_1 > q_{\rm max}/2$ and $m_2 \le q'_{\rm max}/2$ (non-Ising)
% \item $m_1 > l_1+\frac{1}{2}$ and $m_2 < l_2+\frac{1}{2}$ (non-Ising)
\begin{eqnarray}
 \hat{\rm H}_{\rm eff} &=&
   \mathcal{J}_{II} \tilde{I}_1 \tilde{I}_2 
 + \mathcal{K}_{zz} \frac{\tilde{s}_{1z_1}}{\tilde{s}_1} \frac{\tilde{s}_{2z_2}}{\tilde{s}_2}
 + \mathcal{J}_{I+} \tilde{I}_1 \tilde{s}_{2+} 
 + \mathcal{J}_{I-} \tilde{I}_1 \tilde{s}_{2-}.
\label{Eq:H_Lines_C2}
\end{eqnarray}
 \item $m_1 \le q_{\rm max}/2$ and $m_2 > q'_{\rm max}/2$ (non-Ising)
% \item $m_1 < l_1+\frac{1}{2}$ and $m_2 > l_2+\frac{1}{2}$ (non-Ising)
\begin{eqnarray}
 \hat{\rm H}_{\rm eff} &=&
   \mathcal{J}_{II} \tilde{I}_1 \tilde{I}_2 
 + \mathcal{K}_{zz} \frac{\tilde{s}_{1z_1}}{\tilde{s}_1} \frac{\tilde{s}_{2z_2}}{\tilde{s}_2}
 + \mathcal{J}_{+I} \tilde{s}_{1+} \tilde{I}_{2} 
 + \mathcal{J}_{-I} \tilde{s}_{1-} \tilde{I}_{2}.
\label{Eq:H_Lines_C3}
\end{eqnarray}
 \item $m_1 \le q_{\rm max}/2$ and $m_2 \le q'_{\rm max}/2$ (non-Ising)
% \item $m_1 < l_1+\frac{1}{2}$ and $m_2 < l_2+\frac{1}{2}$ (non-Ising)
\begin{eqnarray}
 \hat{\rm H}_{\rm eff} &=&
   \mathcal{J}_{II} \tilde{I}_1 \tilde{I}_2 
 + \mathcal{K}_{zz} \frac{\tilde{s}_{1z_1}}{\tilde{s}_1} \frac{\tilde{s}_{2z_2}}{\tilde{s}_2}
\nonumber\\
 &+&
   \mathcal{J}_{I+} \tilde{I}_1 \tilde{s}_{2+} 
 + \mathcal{J}_{I-} \tilde{I}_1 \tilde{s}_{2-} 
 + \mathcal{J}_{+I} \tilde{s}_{1+} \tilde{I}_{2} 
 + \mathcal{J}_{-I} \tilde{s}_{1-} \tilde{I}_{2} 
\nonumber\\
 &+&
   \mathcal{J}_{++} \tilde{s}_{1+} \tilde{s}_{2+} 
 + \mathcal{J}_{+-} \tilde{s}_{1+} \tilde{s}_{2-} 
\nonumber\\
 &+&
   \mathcal{J}_{-+}  \tilde{s}_{1-} \tilde{s}_{2+} 
 + \mathcal{J}_{--}  \tilde{s}_{1-} \tilde{s}_{2-}.
\label{Eq:H_Lines_C4}
\end{eqnarray}
 \end{enumerate}
\end{enumerate}

\subsection{Doublet and isotropic spin}
The effective Hamiltonian between the anisotropic doublet and isotropic spin (\ref{Eq:H_doub_iso})
reduces as follows:
\begin{enumerate}
  \item $m_1 > q_{\rm max}/2$ (Ising, II)
%  \item $m_1 > l_i + \frac{1}{2}$ (Ising, I)
  \begin{eqnarray}
   \hat{\rm H}_{\rm eff} &=& \sum_{q'=-1}^1 \mathcal{K}_{zq'} 
   \frac{\tilde{s}_{1z_1}}{\tilde{s}_1}
   \frac{{\rm S}_{2q'}}{S_2}
\\
    &=&
   \mathcal{K}'_{zz'} 
   \frac{\tilde{s}_{1z_1}}{\tilde{s}_1}
   \frac{{\rm S}_{2z'}}{S_2},
  \end{eqnarray}
where 
  \begin{eqnarray}
   \mathcal{K}'_{zz'} &=& \sqrt{\sum_{q'=-1}^1 |\mathcal{K}_{zq'}|^2},
\\
   {\rm S}_{2z'} &=& \sum_{q'=-1}^1\frac{\mathcal{K}_{zq'} {\rm S}_{2q'}}{\mathcal{K}'_{zz'}}.
  \end{eqnarray}
  \item $m_1 \le q_{\rm max}/2$ (non-Ising)
%  \item $m_1 \le l_i + \frac{1}{2}$
  \begin{eqnarray}
   \hat{\rm H}_{\rm eff} &=& 
  \sum_{q'=-1}^1 
  \left[
   \mathcal{K}_{zq'} \frac{\tilde{s}_{1z_1}}{\tilde{s}_1}
 + \mathcal{K}_{+q'} \tilde{s}_{1+}
 + \mathcal{K}_{-q'} \tilde{s}_{1-}
  \right]
  \frac{{\rm S}_{2q'}}{S_2}.
 \end{eqnarray}
\end{enumerate}

% Acknowledgment
\ack
N. I. would like to acknowledge the financial support from 
the Fonds Wetenschappelijk Onderzoek - Vlaanderen (FWO) 
%the Flemish Science Foundation (FWO)
and the GOA grant from KU Leuven.

% Reference
\section*{References}
%\bibliographystyle{iopart-num}
%\bibliography{ref}

\providecommand{\newblock}{}

\end{document}